\documentclass[opre,nonblindrev]{informs3_hide}
\DoubleSpacedXI % default for OR
\usepackage[english]{babel}
\usepackage[autostyle, english = american]{csquotes}
\MakeOuterQuote{"}

\usepackage{bbm, xspace}
\usepackage[hypertexnames=false]{hyperref}
\usepackage{cleveref}
\crefname{subsection}{subsection}{subsections}
\usepackage[normalem]{ulem}

\usepackage{caption}
\usepackage{subcaption}

\newcommand{\bE}{\mathbb{E}}

\newcommand{\ALG}{\mathsf{ALG}}
\newcommand{\OPT}{\mathsf{OPT}}
\newcommand{\VO}{\mathsf{VO}}
\newcommand{\EO}{\mathsf{EO}}

\usepackage[dvipsnames]{xcolor}

\usepackage{natbib,bbm,multirow,multicol}
 \bibpunct[, ]{(}{)}{,}{a}{}{,}%
 \def\bibfont{\small}%
\TheoremsNumberedThrough     % Preferred (Theorem 1, Lemma 1, Theorem 2)
\ECRepeatTheorems

\EquationsNumberedThrough    % Default: (1), (2), ...
\MANUSCRIPTNO{OPRE-2023-02-061.R1} % When the article is logged in and DOI assigned to it,
\usepackage{tabstackengine}
\usepackage[nopar]{lipsum}

\usepackage{algorithm}
\usepackage{algpseudocode}

\newcommand{\PP}{\mathbb{P}}
\newcommand{\E}{\mathbb{E}}

\newcommand{\LP}{\textsc{LP}}

\newcommand{\DR}{\textsc{DR}}

\newcommand{\DD}{\mathcal{D}}

\newcommand{\R}{\mathbb{R}}
\newcommand{\I}{\mathcal{I}}

\newcommand{\Ind}{\mathbbm{1}}

\newcommand{\FF}{\mathcal{F}}

\newcommand{\tot}{\mathtt{total}}
\newcommand{\high}{\mathtt{high}}
\newcommand{\seq}{\mathtt{seq}}

\newcommand{\paral}{\mathtt{par}}
\newcommand{\ptk}{\mathtt{ptk}}
\newcommand{\simul}{\mathtt{sim}}

\newcommand{\ALT}{\mathtt{alt}}
\newcommand{\bin}{\text{Bin}}
\newcommand{\poiss}{\mathrm{Pois}}

\newcommand{\orrev}[1]{{#1}}

\begin{document}
\DoubleSpacedXI % default for OR
%%%%%%%%%%%%%%%%
% \linespread{\linespace}\selectfont{}
%\setcounter{page}{1}
%\DOI{10.1287/xxxx.0000.0000}%

\RUNAUTHOR{Epstein and Ma}

\RUNTITLE{Selection and Ordering Policies for Hiring Pipelines}

\TITLE{
% Understanding Hiring through Linear Programming
Selection and Ordering Policies for Hiring Pipelines via Linear Programming
}

\ARTICLEAUTHORS{
\AUTHOR{Boris Epstein}
\AFF{Graduate School of Business, Columbia University, New York, NY 10027, \EMAIL{bepstein25@gsb.columbia.edu}}

\AUTHOR{Will Ma}
\AFF{Graduate School of Business, Columbia University, New York, NY 10027, \EMAIL{wm2428@gsb.columbia.edu}}
}

\ABSTRACT{

Motivated by hiring pipelines, we study three selection and ordering problems in which applicants for a finite set of positions must be interviewed or sent offers. There is a finite time budget for interviewing/sending offers, \orrev{and every interview/offer is followed by a stochastic realization of discovering the applicant's quality or acceptance decision}, leading to computationally challenging problems. In the first problem, we study sequential interviewing and show that a computationally tractable, non-adaptive policy that must make offers immediately after interviewing is \orrev{near-}optimal, assuming offers are always accepted. \orrev{We further show how to use this policy as a subroutine for obtaining a PTAS.} In the second problem, we assume that applicants have already been interviewed but only accept offers with some probability; we develop a computationally tractable policy that makes offers for the different positions in parallel, which can be used even if positions are heterogeneous, and is \orrev{near-}optimal relative to a policy that can make the same \orrev{total number} of offers \orrev{one by one}. In the third problem, we introduce a parsimonious model of overbooking where all offers must be sent simultaneously and a linear penalty is incurred for each acceptance beyond the number of positions; we provide nearly tight bounds on the performance of practically motivated value-ordered policies.

All in all, our paper takes a unified approach to three different hiring problems, based on linear programming.  Our results in the first two problems generalize and improve the \orrev{existing guarantees due to} \citet{purohit2019hiring} that were between 1/8 and 1/2 to new guarantees that are at least $1-1/e\approx 63.2\%$. We also numerically compare three different settings of making offers to candidates (sequentially, in parallel, or simultaneously), providing insight into when a firm should favor each one.

% Motivated by hiring pipelines, we study two order selection problems in which applicants for a finite set of positions must be interviewed or made offers sequentially.  There is a finite time budget for interviewing or making offers, and a stochastic realization after each decision, leading to computationally-challenging problems.  In the first problem we study sequential interviewing, and show that a computationally-tractable, non-adaptive policy that must make offers immediately after interviewing is approximately optimal, assuming offerees always accept their offers.  In the second problem, we assume that applicants have already been interviewed but only accept offers with some probability; we develop a computationally-tractable policy that makes offers for the different positions in parallel, which is approximately optimal even relative to a policy that does not need to make parallel offers.  Our two results both generalize and improve the guarantees in the work of \citet{purohit2019hiring} on hiring algorithms, from 1/2 and 1/4 to approximation factors that are at least $1-1/e\approx 63.2\%$.

}

%\KEYWORDS{}

% \linespread{\linespace}\selectfont{}
\maketitle
%%%%%%%%%%%%%%%%%%%%%%%%%%%%%%%%%%%%%%%%%%%%%%%%%%%%%%%%%%%%%%%%%%%%%%

% \section{Introduction}

% \section{Model and Notation}

% \linespread{\linespace}\selectfont{}

% \textcolor{red}{\textbf{Reminder for Will}: find a reference for the proof of the `folklore splitting argument'.}

\section{Introduction}
%\linespread{\linespace}\selectfont{}

Hiring the right personnel is {one of} the most important factors in the success of an enterprise. That being said, carrying out an efficient and timely recruitment process {can be} challenging in practice. Several difficulties arise when hiring, such as (but not limited to) deciding when to carry out the process, dealing with imperfect information, and making operational decisions at the time that the process is being carried out. This last aspect of the recruitment process raises several questions that can be studied through an algorithmic lens.

% Good recruiting policies are central to the success of any enterprise, yet carrying out efficient and timely recruiting processes can prove to be challenging. Several difficulties arise when hiring personnel, such as deciding when to carry out the process, dealing with imperfect information, or the operation of the process at the time it is carried out, just to name a few. This last aspect raises several questions that can be studied through an algorithmic lens.

A typical recruitment process starts with a firm making a call for applications. An application usually consists of a resume and potential complementary materials. Based on this (imperfect) information, the firm must decide \textit{who} is going to be interviewed and \textit{in which order} are the interviews going to be conducted. Both of these aspects are relevant because a recruitment process cannot go on forever. That is, \orrev{there is not enough time to interview every single applicant, and the firm can choose who to interview next depending on the outcomes of past interviews.}

Applicants become candidates once they are interviewed. After all interviews are conducted, the firm must send offers to the candidates it wishes to hire, given a limited set of positions.
% Again, \textit{who} to send an offer to is the central decision to make here.
However, there are many possible ways to send offers to candidates. The first natural approach would be to \textbf{sequentially} send offers to candidates. By this we mean: send an offer, wait for the response of the candidate, and (if there is still a position available) carry on with the next offer. As in the interviewing process, the \textit{order} in which the offers are sent becomes a relevant decision. A second approach, applicable only to a firm hiring more than one person, is to save time by sending offers in \textbf{parallel}. This means, for each position remaining to fill: send an offer, wait for the response of the candidate, and (if still unfilled) carry on with subsequent offers. The final approach, which may be desirable under a tight timeline or to avoid revealing preferences among the candidates sent offers, is to send all offers \textbf{simultaneously}.
% This last method of sending offers has the benefit that the amount of offers to be sent is not subject to time constraints.
However, it runs the risk of hiring more people than positions available, which must come at a cost.

To answer these operational questions and compare the different modes of sending offers, we study three different models of hiring processes that build off existing work.

\textbf{First model: sequential interviewing, a.k.a.\ ProbeTop-$k$.} In the first model, we assume that the firm has to hire up to $k$ people from a pool of $n$ applicants. Each applicant has a random value unknown to the firm carrying out the hiring process. The firm has access to distributional knowledge of these values coming from the applicants' resumes and complementary material submitted. The firm can interview applicants to find out the realization of their values, but there is a limit of $T$ on the number of interviews conducted. The realization of the value of the applicant becomes known to the firm immediately after carrying out the interview.
We note that this realization should be interpreted as the applicant's \textit{expected} value to the firm given the interview (and conditional on them accepting the offer). \orrev{We treat this value as deterministic, which does not lose generality for a risk-neutral firm that maximizes its expectation.}
We further assume that realizations from interviews are independent across applicants.
% , which is also justifiable in practice.
After all interviews are carried out, the firm can choose the best $k$ interviewed candidates to be hired, who are assumed to accept their offers with probability 1. The goal of the firm is to maximize the expected sum of values of the hired personnel. This problem is exactly the ProbeTop-$k$ problem, as described in \citet{fu2018ptas}.

% which is computationally challenging because the state space is exponential in the number of applicants (one has to track the subset that has been interviewed and their values).

In this problem, we can further distinguish between different classes of policies. First, we distinguish between \textit{adaptive} and \textit{non-adaptive} policies. Adaptive policies can decide the order of the interviews on the fly, choosing who to interview next depending on the outcomes of previous interviews. Non-adaptive policies, in contrast, have to fix an interview order before the process starts, which although restrictive, is attractive from an {ease-of-implementation} perspective. We also distinguish between \textit{committed} and \textit{non-committed} policies. Committed policies have to irrevocably decide whether to hire each candidate immediately after interviewing them and discovering their value. Non-committed policies, in contrast, can carry out all interviews and choose the $k$ highest realized values in hindsight. Using committed policies could be attractive from a practical point of view, as waiting until the end incurs the risk that candidates accept offers from competing firms in the meantime. 
We are interested
% both in approximately-optimal algorithms and
in bounds on how costly it is to restrict the firm to use policies that are non-adaptive and committed. \orrev{The former is quantified in the literature by the widely studied notion of \textit{adaptivity gap}: the worst-case ratio between the performance of general policies vs.\ algorithms that are restricted to be non-adaptive.
% We also introduce the \textit{commitment gap}, the analogue ratio but when restricting to committed policies.
Our results will bound the "adaptivity-commitment gap", in which the algorithm is restricted to be both non-adaptive and committed.}

\textit{Relation to Free-Order Prophets.}
The Free-Order Prophet Inequality problem is the special case of ProbeTop-$k$ where $T=n$ and 
only committed policies are allowed. (When $T=n$, the constraint of $T$ interviews is not binding, and hence non-committed policies can just trivially interview all applicants.)
Typically in prophet inequalities, the benchmark can see all applicants' values {in advance and simply choose to interview and hire the $k$ overall highest-valued candidates.}
%and does not need to commit (i.e.\ they can hire the $k$ highest-valued candidates in hindsight).
We note that such a benchmark is \textit{too powerful to compare against} in our more general problem if $n$ is much larger than $T$ since the benchmark sees all $n$ realizations while the algorithm can only interview $T$ applicants.
This is why in our general ProbeTop-$k$ problem, we compare to an optimal (adaptive, non-committed) algorithm that is still bound by $T$ interviews that must be decided without any prophetic information, making our comparison different from prophet inequalities.

% However, our algorithm for ProbeTop-$k$, being non-committed, implies an algorithm for free-order prophets that has the same guarantee relative to the prophet benchmark in the special case of free-order prophet inequalities (where $T=n$). \wnote{Let's just remove this sentence.  We don't actually formalize that our benchmark captures the prophet benchmark when $T=n$ right?  Also, the $1-e^{-k}k^k/k!$ result for free-order prophets is already known to Yan (2011).}

\textit{Relation to Sequential Offering.}
% Our sequential interviewing model
% is closely related to
In the Sequential Offering model of \citet{purohit2019hiring},
applicants are assumed to have already been interviewed but have uncertainty about whether they accept an offer.
The firm knows, for each candidate, how likely it is that they will accept an offer, and the value they add to the firm, should they accept.
The firm has time to send at most $T$ offers and wants to maximize the expected total value of up to $k$ candidates who accept their offers.
%We show that this setting can be captured by 
\orrev{This setting is closely related to the special case of ProbeTop-$k$ with weighted Bernoulli distributions, where the values take a positive realization with some probability (representing an acceptance) or 0 with the remaining probability (representing a rejection). The subtle difference is that an accepted offer cannot be withdrawn by the firm in the Sequential Offering model, whereas in the ProbeTop-$k$ model, the firm can turn down a candidate even if their value turns out to be positive. We will show that our algorithm satisfies properties that makes it admissible for this Sequential Offering problem too, and improve upon the results of \citet{purohit2019hiring}.}

\textbf{Second model: Parallel Offering.}
We also study a Parallel Offering model, in which
% The second model we study is the parallel offering problem, an extension of the sequential offering problem studied by \citet{purohit2019hiring}.
a firm again has to hire people to fill $k$ positions. However, we now allow for heterogeneous positions, where a candidate may have different potential values for different job positions. We assume that all interviews have already been conducted, leaving us with a pool of $n$ desirable candidates. After conducting all interviews the firm learned, for each candidate, how valuable they are for each of the available positions and how likely it is for each candidate to accept an offer for each of the available positions. The firm must now decide how to send offers in $T$ parallel offering rounds. At each round, the firm can send an offer for each of the positions that have not yet been filled by a candidate. When a candidate receives an offer, they can either accept or reject it, with the assumption that they cannot receive an offer {for another position} if they rejects it (and hence candidates do not try to anticipate offers they might receive later). The goal of the firm is to maximize the expected sum of values of hired candidates. We develop a non-adaptive algorithm that can be computed efficiently and performs competitively in comparison to adaptive and even relaxed sequential algorithms.

% \textbf{Relation to parallel offering model of \cite{purohit2019hiring}.}
This model generalizes a parallel offering model also introduced in \citet{purohit2019hiring}, which is similar but has $k$ identical instead of heterogeneous positions.
% , justifying the assumption that each candidate can only be offered once.
Our Parallel Offering model is not only more general, but we also derive stronger performance guarantees.
\strutlongstacks{T}

\textbf{Third model: Simultaneous Offering.}
Finally, we study the Simultaneous Offering model, again for a firm hiring to fill $k$ positions.
% The third and final model that we study is the simultaneous offering model. As in the previous two models, we assume that the firm has to hire people in order to fill $k$ positions.
As in the Parallel Offering model, the firm has already conducted all interviews, resulting in a pool of $n$ candidates for which it knows how valuable each candidate is to the firm and how likely each candidate is to accept an offer. The firm must decide on a subset of candidates who will receive an offer (all at the same time). This subset {can be of any size}, so a possible outcome is that more than $k$ candidates accept an offer. If that is the case, the firm must pay a penalty for each candidate who accepted an offer beyond the capacity $k$. This penalty can be thought of as the cost of withdrawing an offer, or the cost of creating a new position in the firm. We analyze the performance of value-ordered policies, which send offers to candidates above a value threshold (regardless of their probability of acceptance). We derive near-optimal approximation guarantees for these policies.

\Cref{tab:intro} contains a comparison of all the models studied/captured in this paper.%\vspace{-35pt}
\SingleSpacedXI
% \linespread{1.1}\selectfont{}
\begin{table}[h] 
\small
\begin{tabular}{|p{.12\textwidth}|p{.12\textwidth}|p{.21\textwidth}|p{.16\textwidth}|p{.2\textwidth}|p{.11\textwidth}|}
\hline 
Model                 & Action \newline performed & Result of action                  & Actions per time step          & Moment of\newline hiring                   & Bound on\newline total hires \\ \hline
ProbeTop-$k$          & Interview        & Observe value of\newline candidate        & One                            & After last interview               & Hard                      \\\hline
Free-Order Prophets   & Interview       & Observe value of\newline candidate        & One                            & After each interview (irrevocable) & Hard                      \\\hline
Sequential\newline Offering   & Send offer       & Observe accept/reject decision    & One                            & Upon acceptance of offer           & Hard                      \\ \hline
Parallel\newline Offering     & Send offer(s)    & Observe accept/reject decision(s) & One per position\newline remaining & Upon acceptance of offer           & Hard                      \\ \hline
Simultaneous Offering & Send offer(s)    & Observe accept/reject decision(s) & $n$                            & Upon acceptance of offer           & Soft (linear penalty) \\ \hline
\end{tabular}
\caption{Comparison between models.} %Each model comes with $n$ applicants/candidates, $k$ positions to fill, and $T$ time steps to perform actions (in Simultaneous offering, $T=1$). The Free-order prophets problem typically is additionally constrained to have $T=n$, but our results hold without this assumption}.
\label{tab:intro}
% \setstretch{2}
% \OneAndAHalfSpacedXI
\end{table}
\DoubleSpacedXI % default for OR
% \linespread{\linespace}\selectfont{}
% \vspace{-35pt}
% \DoubleSpacedXI

\subsection{\orrev{Outline of Results }\label{sec:results_proofs}}

% \wnote{Maybe change to just "Results" or "Outline of Results"?}

\orrev{We will generally say that our algorithm is $\alpha$-approximate if its expected total value collected is at least $\alpha$ times that of an optimal algorithm, from a larger class.  We call $\alpha\in[0,1]$ the approximation factor.
We note that typically this terminology is used when comparing to the optimal algorithm from the \textit{same} class.
Our results imply lower bounds on the approximation factor $\alpha$ under the typical terminology, since we are comparing against a larger class.
All of our results assume that distributions have \textit{finite support} and are explicitly input in the form of (value, probability) pairs
with binary encoding.
All of our algorithms are polynomial-time under this form of input.}

\orrev{
We should note that for the problem of computing the optimal algorithm within a fixed class, no computational hardness results are known for any of the problems we study (ProbeTop-$k$, Parallel Offering, Simultaneous Offering), to our knowledge.
Nonetheless, our algorithms still have some of the currently best-known approximation factors, which also hold when comparing to a larger class.}

% For $\alpha\in [0,1]$, we generally say that an algorithm is  from a larger class.
% \orrev{We call} $\alpha$ the \textit{approximation factor}. \orrev{It is worth noting that there are no concrete hardness results for any of the problems we study (ProbeTop-$k$, Parallel Offering, and Simultaneous Offering), but there is still a prominent branch of literature dedicated to finding approximation algorithms for them.}

\orrev{\textbf{Sequential interviewing a.k.a.\ ProbeTop-$k$ problem.}
We develop a polynomial-time algorithm that is non-adaptive, committed, and achieves a $(1-e^{-k}k^k/k!)$ approximation factor relative to an optimal adaptive, non-committed algorithm
% \orrev{when the distributions of applicants' valuations have \emph{finite support}}
(\Cref{thm:probetopk}, \Cref{sec:ptk_analysis}).  We note that the approximation factor of $(1-e^{-k}k^k/k!)$ equals $1-O(1/\sqrt{k})$ by Stirling's approximation, is always at least $1-1/e\approx0.632$ (when $k=1$), and increases to 100\% as $k\to\infty$.  We also note that it is tight relative to the LP benchmark we compare against (\Cref{sec:tight}).}

\orrev{Our results for this problem, while simple and clean, have broad implications.
First, we can combine our algorithm with the work of \citet{fu2018ptas} to obtain a PTAS for ProbeTop-$k$ (\Cref{cor:ptas}, \Cref{sec:ptas}). There exist PTAS's for ProbeTop-$k$ restricted to non-adaptive algorithms \citep{segev2021efficient} and ProbeTop-$k$ restricted to committed algorithms \citep{fu2018ptas}, but a PTAS for the general (adaptive, non-committed) ProbeTop-$k$ problem appears unknown, assuming that $k$ is part of the input. Second, the best-known existing guarantee on the adaptivity gap for ProbeTop-$k$ was 1/2, due to \citet{bradac2019near} via an algorithm that is not necessarily polynomial-time. We improve the lower bound on the adaptivity gap for ProbeTop-$k$ from $1/2$ to $1-1/e$ using a non-adaptive, polynomial-time algorithm, and moreover show asymptotic optimality when the number of positions grows to infinity. Third, our algorithm can be directly applied to the Sequential Offering problem without a loss (\Cref{cor:seq}, \Cref{subsec:weighted_ber}), where the best-known approximation factor and adaptivity gaps were $1/2$ \citep{purohit2019hiring}. As with \citet{bradac2019near}, the adaptivity gap shown in \citet{purohit2019hiring} was achieved through a non-constructive approach.}
%First, the previously best-known approximation factor was 1/2, obtained by both \citet{purohit2019hiring} and \citet{bradac2019near}, so our lower bound is not only stronger for all values of $k$, but demonstrates asymptotic optimality in $k$.
%Second, the $1/2$-approximate algorithm in \citet{bradac2019near} is not polynomial-time and not committed, while the polynomial-time $1/2$-approximate algorithm in \citet{purohit2019hiring} needs to be adaptive, only compares against policies that are committed, and also focuses on the special case of sequential offering (we elaborate on this in \Cref{subsec:weighted_ber}).
Finally, our abstract treatment leads to an algorithm that works in the Free-Order Prophets setting (because it is committed), hence our results extend to a free-order "prophet inequality" (comparing against a necessarily weaker benchmark) that allows an additional constraint of $T$ on the number of interviews.

\textbf{Parallel Offering problem.}
For the Parallel Offering model, we develop an algorithm that is non-adaptive and achieves a $(1-1/e)$ approximation factor relative to an optimal (adaptive) algorithm (\Cref{thm:paral}, {\Cref{sec:par_analysis}}).
This result also both improves and generalizes existing results, in this case, the 1/8-approximate algorithm of 
%This result improves upon 1/4, the best known bound known for this problem, derived by
\citet{purohit2019hiring} that works in the special case where all positions are identical.
It is also tight relative to the LP benchmark we compare against (\Cref{sec:parTight}). We further show that our algorithm obtains at least $(1-1/e)$ times what an optimal algorithm would obtain in the Sequential Offering problem with $kT$ time steps, establishing a lower bound on the value of batching offers (\Cref{cor:cost_of_batching}, \Cref{sec:batching}).

\textbf{Simultaneous Offering problem.} We analyze the performance of \textit{value-ordered} policies, which send offers to {all candidates whose value is above a threshold}.
% This means that if the policy wants to send an offer to the candidate with the $m$-th highest value, then it is obliged to send an offer to the $m-1$ candidates with a higher value as well.
The idea of having a value threshold is practically motivated and the optimal value-ordered policy can be computed efficiently. We show that with no assumptions on the values of the candidates, value-ordered policies can have arbitrarily poor performance (\Cref{eg:simul_VO}, \Cref{sim:policies}). \orrev{Consequently, we assume that the valuations of candidates are lower bounded by a parameter $\tau\in(0,1)$, and provide a value-ordered policy that achieves at least a factor $\alpha_k^\tau$ of what an optimal policy could achieve, {where $\alpha_k^\tau$ is an increasing function of $k$ and $\tau$} (\Cref{thm:simul_lower_bound}, \Cref{sec:simLB}) {that satisfies $\lim_{k_\to\infty} \alpha_k^\tau = 1$ for every $\tau\in(0,1)$ (\Cref{thm:simul_asymptotic}, \Cref{sec:simLB})}. We also provide an instance where no value-ordered policy can achieve a factor higher than $\beta_k^\tau$ of what an optimal policy could achieve, {where $\beta_k^\tau$ is another increasing function of $k$ and $\tau$} (\Cref{thm:simul_upper_bound}, \Cref{sec:simLB}). \orrev{We further} characterize the region where $\alpha_k^\tau = \beta_k^\tau$ (\Cref{prop:simul_tight}, \Cref{sec:simLB}). In particular, our bound is tight for all $k$ if $\tau \leq 1/2$.}

\orrev{\textbf{Numerical study.}} In \Cref{sec:numerics}, we numerically simulate the three different models of sending offers (sequential, parallel, simultaneous) using the same candidate pool generation process as \citet{purohit2019hiring}.
We illustrate the improved performances offered by our new policies, and compare the performances attainable across the different models, providing insight into when the firm should favor sending offers sequentially, in parallel, or simultaneously. For instance, there is surprisingly little value to gain from an algorithm that sends offers in parallel, as opposed to sending single offers sequentially and being highly adaptive to the number of remaining positions, unless the horizon for making offers is extremely short.
% In the same vein, when the amount of time steps available is large, a firm is better off sending offers sequentially than sending offers in parallel in a non-adaptive fashion, even if the sequential policy is also non-adaptive.
For Simultaneous Offering, we demonstrate that value-ordered policies are generally desirable unless there is both a small number of initial positions and a high cost of overage.  In that case, it is better to identify the highest-valued "safe" candidates (with a high probability of acceptance) to reduce the variance in the number of accepted offers. \orrev{We acknowledge that these insights are based on the specific generative model of \citet{purohit2019hiring} and they do not necessarily hold in general.} 
% The simultaneous offering paradigm is more valuable to a firm when there is a large amount of positions to be filled and a low amount of time steps to send offers. In this regime, simultaneous offering policies outperform sequential policies even when the cost of hiring over capacity is large relative to the value of the candidates.
% \wnote{I tried to highlight the insights that were more "surprising"; uncomment my changes if you agree.}

\section{Related Work \label{sec:related_work}}

% In this section, we provide an overview of our results and their connection to the existing literature. In \Cref{sec:results_proofs} we give a summary of our results and their relevance with respect to existing work.  In \Cref{sec:related_work} we provide a broader list of literature that surrounds our work.

% \subsection{Further Related Work \label{sec:related_work}}

% \orrev{In this section, we provide a broad discussion about the literature that surrounds our work.}
% \wnote{Do we need this sentence?  Also, maybe call it Related Work since we are not trying to be so exhaustive.}

\textbf{Stochastic Probing and Matching.} Our work is closely related to the general stochastic probing problem studied in \citet{gupta2013stochastic} and \citet{gupta2016algorithms}. These papers study the problem of sequentially probing elements in order to maximize the sum of the weights of a selected subset. In their setting they consider more general sets of `outer' constraints to be satisfied by the probed elements, and `inner' constraints to be satisfied by the selected elements. In this language, ProbeTop-$k$ considers the outer constraint to be the $T$-uniform matroid and the inner constraint to be the $k$-uniform matroid. Their work is different from ours in that it only allows the probe to have two outcomes (active or inactive) and active probes must be irrevocably included in the final subset. \citet{bradac2019near} introduce the multiple-type general stochastic probing problem. In this work, they only work with outer constraints and include the inner constraints by allowing submodular functions instead of modular functions.

In \citet{bradac2019near} they show that the adaptivity gap
% \footnote{This is defined as the worst-case ratio between the performance of adaptive and non-adaptive policies. \wnote{We already defined this earlier?  Can remove footnote.}}
is exactly 1/2 for the stochastic probing problem with monotone submodular functions under prefix-closed probing constraints. Their proof is not constructive, in the sense that the algorithm for their lower bound requires the optimal decision tree as an input. The best-known non-adaptive algorithm for which is known that it can be computed in polynomial time is due by \citet{gupta2017adaptivity}, which achieves a 1/3 approximation for submodular and XOS functions under prefix-closed probing constraints.

We note that these stochastic probing problems were heavily inspired by the stochastic matching problem with patience constraints, originally studied in \citet{chen2009approximating} and \citet{bansal2012lp}.  Our Parallel Offering problem has the flavor of a stochastic matching problem, although it is heavily constrained and simpler.

\textbf{PTAS-type results.}
\orrev{\citet{fu2018ptas} develop PTAS's for a class of dynamic programs, that includes ProbeTop-1 over general (adaptive, non-committed) algorithms, and ProbeTop-$k$ over committed algorithms.
They also provide a $(1-\varepsilon)-$approximation for ProbeTop-$k$ over general algorithms whose runtime is polynomial in $1/\varepsilon$ but exponential in $k$.
\citet{segev2021efficient} develop the improved notion of EPTAS's for several related problems, including ProbeTop-$k$ over non-adaptive algorithms.
Their EPTAS works differently when $k$ is small or large. We use the same LP relaxation as theirs for the large $k$ case, although our rounding scheme and analysis are very different.} We note that none of these PTAS-type results use non-adaptive, committed policies to approximate the best adaptive, non-committed policies, as in our paper.

\textbf{Prophet Inequalities.}
Our work has the flavor of Prophet Inequalities in that we are deciding online whether to accept values drawn from known distributions, and comparing against a supernatural benchmark.
Classical works in Prophet Inequalities \citep{krengel1977semiamarts,krengel1978semiamarts,samuel1984comparison,kleinberg2012matroid,esfandiari2017prophet} assume that the order is beyond our control, while we are studying the \textit{free-order} variant where the order can be decided \citep{hill1983prophet,beyhaghi2020improved,agrawal2020optimal}.
However, due to the constraint of $T$ time steps, as discussed earlier, it is important {to emphasize} that the benchmark we are comparing against is also bound by $T$ time steps, so we are not comparing against (and it is impossible to compare against) the true prophet.
For sequential interviewing, the approximation factor we derive of $(1-e^{-k}k^k/k!)$ has also been established for free-order by \citet{yan2011mechanism} and extended in \citet{arnosti2021tight} to random-order, but neither of these works allows for a constraint of $T$ time steps.

\textbf{Simultaneous Offering model.} Our Simultaneous Offering model and in particular the linear penalty cost is motivated by static overbooking models in revenue management \citep[see e.g.][Ch.~3]{gallego2019revenue}.
However, our decision is different in that we are selecting a \textit{subset} of candidates, whereas they are setting a single booking limit (possibly one for each fare class).
The simple family of heuristics for which we provide an approximation guarantee also has no analog in overbooking. Our problem also shares many aspects with the one studied in \citet{cominetti2010optimal}, where a subset of customers are sent last-minute offers for a limited amount of items. Their model differs from ours in that if more items than available inventory are sold, they only collect value from a random subset of size $k$; whereas, our model collects the value of all acceptances but has to pay a linear penalty.
% Our offline subset selection problem is actually a special case of (non-monotone) submodular maximization, for which there exists a 1/2-approximation using a complicated algorithm \wnote{add reference}.
% By exploiting the special structure in our problem in certain regimes, we achieve improved approximations via the much simpler family of value-ordered policies.

\textbf{Concurrent work.}
Parallel to our work, \citet{gallego22aconstructive} have studied and obtained results for the ProbeTop-$1$ problem. Although both works aim to derive non-adaptive algorithms for the problem, theirs is different from ours in several aspects. On one hand, their models assume that the valuations of the candidates come from independent random variables with either general distributions or continuous distributions, whereas our work focuses on random variables with finite support. On the other hand, the bounds they achieve are 1/2 for general random variables and $1-1/e$ for continuous random variables, which also hold for the ProbeTop-$k$ problem, but they do not improve as $k$ grows. Our bounds, in contrast, grow from $1-1/e$ when $k=1$, to 1 when $k\to\infty$. One last aspect that distinguishes our work from \cite{gallego22aconstructive} is the relaxations used to measure the performance of their algorithms. While we use standard LP relaxations, they introduce a novel benchmark consisting of a simple minimax problem.

\section{Problem Formulations and Preliminaries}

In this section, we formally state the problems studied in the paper.
% \wedit{We assume throughout the formulations that all probability distributions are finitely-supported.} 
We first state the ProbeTop-$k$ problem in \Cref{sec:ptk_intro}. We then state the Parallel Offering problem in \Cref{sec:parintro}. We close by stating the Simultaneous Offering problem in \Cref{sec:simul_intro}.

%We close with Section \ref{sec:known_results}, where we mention some results in the literature that will be useful for both constructing our algorithms and carrying out our analyses. \{add simul to roadmap}

\subsection{ProbeTop-k Problem \label{sec:ptk_intro}}

In the ProbeTop-$k$ ($\ptk$) problem, a firm faces the challenge of filling $k$ positions out of a pool of $n$ applicants. Each applicant $i$ has a random, non-negative value $V_i\sim F_i$, and $F_i$ is known to the firm. Before the firm hires an applicant, an interview must be conducted. When the firm interviews applicant $i$, the realization of $V_i$ becomes known to the firm. The firm can conduct at most $T$ interviews in total and then can choose any $k$ interviewed candidates to be hired. The goal of the firm is to maximize the sum of the values of the hired candidates. An instance of the problem is characterized by a tuple $I=(k,T,n,F)$, where $1\leq k\leq T \leq n$, and $F = \{F_i\}_{i\in [n]}$
%\footnote{We use $[n]$ to denote $\{1,\dots,n\}$.}
is a collection of probability distribution functions. In this work, we focus on distributions supported on a finite set {of non-negative values} $\{r_j\}_{j\in [J]}$ %\orrev{with} polynomial size
and we use $q_{ij}$ to denote $\PP(V_i = r_j)$. We use $\I_\ptk$ to denote the set of all possible instances for the ProbeTop-$k$ problem. We further use 
$\I_\ptk^k \subseteq \I_\ptk$ to denote the subset of instances where the number of positions is k.
% to denote $\{I\in\I_\ptk:$ amount of positions is $k\}$.

\orrev{For this problem, we define a policy as a function $\pi$ that maps the remaining budget of interviews, the set of applicants that have not yet been interviewed, and the realization of the values of the candidates that have already been interviewed to a decision of which applicant to interview next.
% \footnote{\orrev{The policy of course takes $k$ into account, but since $k$ is fixed during the instance, we take it as given.}}
}
Let $\Pi^\ptk$ be the set of all policies for $\ptk$. For a policy $\pi\in\Pi^\ptk$ and an instance $I\in\I_\ptk$, let $R_\pi(I)$ be the expected reward of using policy $\pi$ on the instance in question. For an instance $I\in\I_\ptk$ define $\OPT_\ptk(I):=\sup_{\pi\in\Pi^\ptk} R_\pi(I)$ as the expected reward of using the best possible policy on instance $I$. We call a policy $\pi \in \Pi^\ptk$ an $\alpha$-approximation if 
\[\inf_{I\in \I_\ptk} \frac{R_\pi(I)}{\OPT_\ptk(I)} \geq \alpha.\] 

We distinguish between \textit{adaptive} and \textit{non-adaptive policies}. A non-adaptive policy has to decide an order in which to conduct the interviews before the process starts. An adaptive policy, in contrast, conducts the interviews sequentially and can use the outcomes of previous interviews to decide which candidate to interview next. We also distinguish between \textit{committed} and \textit{non-committed} policies. After each interview, committed policies must irrevocably decide whether to hire the candidate or not. Non-committed policies, in contrast, can interview $T$ applicants and then choose the $k$ highest realizations among them. {Among other questions,} we are interested in how well can the firm perform when restricted to using non-adaptive policies, committed policies, or both. To quantify this, we are interested in how good an approximation factor can be achieved by restricting the firm to using policies that are both non-adaptive and committed.

\subsection{Parallel Offering Problem \label{sec:parintro}} 

The second setting we study in this paper is the Parallel Offering problem ($\paral$). Again consider the case of a firm that has to hire candidates in order to fill $k$ positions. Instead of conducting interviews, the firm has to send offers to candidates. The positions of the firm are now not identical, and for each candidate $i\in[n]$ and position $j\in[k]$ the firm knows $v_{ij}$: the reward collected by the firm if candidate $i$ accepts an offer for position $j$, and $p_{ij}$: the probability that the candidate $i$ accepts an offer for position $j$. The firm can carry out at most $T$ offering rounds. At each round, the firm can send an offer for each unfilled position in parallel. Each candidate can receive at most one offer in total. The goal of the firm is to maximize the expectation of the sum of the values of the accepted offers. An instance for this problem is defined by a tuple $I=(k,T,n,p,v)$, where $1\leq k\leq T \leq n$, $p\in[0,1]^{n\times k}$ and $v\in \R_+^{n\times k}$. \orrev{For this problem, we define a policy as a function $\pi$ that maps the remaining number of time steps, the remaining set of unfilled positions, and the set of candidates that have not yet received an offer, to an assignment of a subset of not-yet-offered candidates to unfilled positions.} Let $\I_{\paral}$ be the set of all instances and $\Pi^\paral$ be the set of all policies for the Parallel Offering model. Again let $R_\pi(I)$ denote the expected reward of using policy $\pi \in \Pi^\paral$ on instance $I \in \I_\paral$. Let $\OPT_\paral(I):=\sup_{\pi\in\Pi^\paral} R_\pi(I)$ be the best possible expected reward that can be obtained from an instance $I\in \I_\paral$. We say that a policy $\pi\in\Pi^\paral$ is an $\alpha$-approximation if 
\[\inf_{I\in \I_\paral} \frac{R_\pi(I)}{\OPT_\paral(I)}\geq \alpha.\] 

In this model, we assume that before the offering rounds start, each candidate has already decided which offers they would accept. We allow for these decisions to be correlated across different positions for a single candidate, but we assume independence across candidates.

\orrev{In this context, we define a non-adaptive policy as follows. For each position, a list of candidates is constructed in a way such that a candidate cannot appear in more than one list, and that each list contains no more than $T$ candidates. Each list must also have a specified fixed ordering of its candidates. For each position, offers are sequentially sent to the candidates in their corresponding list, in its corresponding order, until either one of the candidates accepts or the list ends without any acceptance (independent of what happens with the other positions). The algorithm we develop for this problem falls under this definition, and thus we establish a lower bound on how well non-adaptive policies can perform.}

%\textcolor{red}{This is the unclear paragraph of comment 43.} In this context, it is not clear how to define non-adaptive policies. A natural class of policies that could be considered non-adaptive are the ones that construct one list with a fixed order for each position and runs the lists in parallel. 

\subsection{Simultaneous Offering Problem\label{sec:simul_intro}}

The third and last setting we study is the Simultaneous Offering problem ($\simul$). As in the previous models we study, a firm faces the challenge of filling $k$ positions out of a pool of $n$ candidates. There are two aspects that make this problem different from $\ptk$ and $\paral$. First, all offers must be made at the same time, and only once (hence the name Simultaneous Offering). The second aspect making this problem different is that the firm can end up hiring over capacity, but pays a linear cost for each candidate hired beyond $k$.\footnote{Requiring the firm to hire at most $k$ candidates with probability 1 would make this problem trivial: the firm would send an offer to the $k$ candidates with the highest expected value $v_i p_i$.}

Formally, we have the following problem. There are $n$ candidates. For each candidate $i\in [n]$, we know $p_i$: the probability that candidate $i$ accepts an offer if sent, and $v_i$: the value obtained if $i$ accepts the offer. The firm must select a subset $S \subseteq [n]$ of candidates to send offers to, which has no cardinality constraint. When the subset $S$ is decided, an offer is sent to each of the candidates in $S$. Each candidate accepts the offer independently with probability $p_i$, in which case they add value $v_i$ to the firm. For each candidate that exceeds the capacity $k$, the firm incurs a linear cost of $c>0$. Let $A\subseteq S$ be the (random) subset of candidates that accepted their offers. The reward obtained by the firm is {$\sum_{i\in A} v_i - c\cdot [|A|-k]^+$}. By re-scaling values $v_i$ by $1/c$, we can assume without loss of generality that $c=1$. The re-scaled values $v_i$ can be interpreted as the value that a candidate adds to the firm, relative to the cost of hiring a candidate over capacity.

With this notation, an instance of $\simul$ can be described by a tuple $I = (k,n,p,v)$, where $1\leq k \leq n$, $p\in[0,1]^n$ and $v\in \R_+^n$. \orrev{We define a policy as a function $\pi$ that maps the problem instance $I$ to a (possibly random) subset of candidates $S\subseteq [n]$.} Let $\I_\simul$ denote the set of all instances and $\Pi^\simul$ denote the set of all policies for the Simultaneous Offering problem. Let $R_\pi(I)$ denote the expected reward of using policy $\pi\in \Pi^\simul$ on instance $I \in \I_\simul$. Let $\OPT_\simul(I) $ be the best possible expected reward that can be obtained from an instance $I \in \I_\simul$.
% \begin{align*}
%     OPT_\simul(I) = \max_{S\subseteq[n] } \quad \E_{A\subseteq S}\left( \sum_{i\in A} v_i - [|A|-k]^+ \right)=\max_{S\subseeqt[n] } \quad  \sum_{i\in S} v_i p_i - \E_{A\subseteq S}\left([|A|-k]^+ \right).
% \end{align*}
We call a policy $\pi$ an $\alpha$-approximation if 
\[ \inf_{I\in \I_\simul} \frac{R_{\pi}(I)}{\OPT_\simul(I)} \geq \alpha.\]
\orrev{One thing worth noting is that, although the objective function of the problem is submodular, it is not monotone and can take negative values. Thus, we cannot use general results for approximating maximization problems with submodular objective functions.}

\section{ProbeTop-k Problem \label{sec:ptk}}

In this section, we provide our algorithm and analysis for the ProbeTop-$k$ problem, where applicants are sequentially interviewed.  We develop a non-adaptive, committed policy that achieves a $(1 - e^{-k}k^k/k!)$ approximation of any (adaptive, non-committed) policy. The algorithm first solves an LP relaxation that upper-bounds the performance of any (adaptive, non-committed) policy, and uses then the LP solution as an input to decide the applicants that will be interviewed, in which order, and whether to hire an interviewed applicant, given their value.

\orrev{In \Cref{sec:ptkLP} we state the LP in question. In \Cref{sec:ptkDR} we introduce a simple dependent rounding scheme that will be used in our approximation algorithm. In \Cref{sec:ptkstatement,sec:ptk_analysis} we introduce and analyze our approximation algorithm. %and in \Cref{sec:ptk_analysis} we show that it attains the mentioned approximation guarantee.
\orrev{In \Cref{sec:ptas} we explain how to use our approximation algorithm as a subroutine for obtaining a PTAS.} In \Cref{subsec:weighted_ber} we treat the special case where values are distributed as weighted Bernoulli random variables \orrev{and extend our results to the Sequential Offering problem}. The analyses in this section, as well as those in \Cref{sec:paral} make use of correlation gap results by \citet{yan2011mechanism} and dependent rounding schemes by \citet{gandhi2006dependent}. We recommend readers who are not familiar with these papers to visit \Cref{sec:known_results}\orrev{, which contains a summary of the relevant technical results contained in the previously mentioned papers,} before reading the proofs. Throughout this section (except for \Cref{subsec:weighted_ber}) we use the word applicant indistinctly from candidate to avoid any confusion, although we had previously distinguished that an applicant becomes a candidate after being interviewed.}

\subsection{Linear Program \label{sec:ptkLP}}

Consider the linear program $\LP_{\ptk}$, which for any instance of ProbeTop-$k$, upper-bounds the performance of all possible policies. \orrev{We acknowledge that linear programs in this spirit have been used exhaustively in the literature. Indeed, this relaxation is used to approach the same problem in \citet{segev2021efficient}, and it consists of a special case of the linear program used in \citet{gupta2013stochastic}.} In $\LP_{\ptk}$, variable $y_i$ is to be interpreted as the probability that applicant $i$ is interviewed. The variable $x_{ij}$ is to be interpreted as the probability that $i$ is hired when $V_i = r_j$.
\begin{align}
    LP_{\ptk}(I)=\max \quad & \sum_{i=1}^n \sum_{j=1}^J r_j x_{ij} \nonumber\\
    s.t. \quad & x_{ij} \leq y_i q_{ij} & \forall i \in [n], j \in [J] \label{const:xyq}  \\
    & \sum_{i=1}^n y_i \leq T & \label{const:yleqT}\\
    & \sum_{i=1}^n \sum_{j=1}^J x_{ij} \leq k& \label{const: xleqk}\\
    & x_{ij}\geq 0 & \forall i \in [n], j \in [J] \nonumber\\
    & 0 \leq y_i \leq 1 & \forall i\in [n]. \nonumber
\end{align}
We first show that $\LP_{\ptk}$ upper-bounds the optimal (adaptive, non-committed) algorithm.

\begin{lemma}
\label{lem:upper_bound_PM}
For any instance $I\in \I_{\ptk}$, we have $\LP_{\ptk}(I)\geq \OPT_\ptk(I)$.
\end{lemma}
{Although this result is not new, we provide a proof of the claim in \Cref{sec:upper_bound_PM_pf} for completeness.}

The following lemma establishes a convenient fact about the basic feasible solutions of the LP. This will let us develop a simple rounding scheme for selecting applicants to be interviewed.
\begin{lemma}\label{lem:bfs2}
Let $I\in\I_\ptk$ and let $y=(y_i)_{i\in[n]},x=(x_{ij})_{i\in[n],j\in[J]}$ be a basic feasible solution of $\LP_\ptk(I)$. Then $y$ has at most 2 non-integer components. If it has 2 non-integer components, they sum to 1.
\end{lemma}
{The proof of this lemma is deferred to \Cref{sec:bfs_pf}.}

\subsection{Dependent Rounding \label{sec:ptkDR}}

We develop a simple dependent rounding scheme (which we call $\DR$ for short) that will be used as a subroutine of our approximation algorithm. $\DR$ receives an optimal solution $y=(y)_{i\in[n]}$ of $\LP_{\ptk}$ as an input and returns a (possibly random) vector $Y\in\{0,1\}^n$. $\DR$ works differently depending on the number of fractional components of the optimal solution $y$. In any case, for all $i$ such that $y_i\in\{0,1\}$, it sets $Y_i = y_i$ with probability 1. If $y$ is integral, then $Y$ is deterministic. If there is exactly one fractional component $i'$, then it sets $Y_{i'}=1$ with probability $y_{i'}$ (and it sets $Y_{i'}=0$ otherwise). If there are exactly two fractional components $i_1$ and $i_2$, then it sets $Y_{i_1}=1,Y_{i_2}=0$ with probability $y_{i_1}$ and sets $Y_{i_1}=0,Y_{i_2}=1$ with probability $1-y_{i_1} = y_{i_2}$.
It is easy to see that the output of $\DR$ satisfies the following two properties: (P1) $\E(Y_i)=y_i$ for every $i\in[n]$, and (P2) $\sum_{i\in[n]} Y_i \leq T$ with probability 1.
% \begin{itemize}
%     \item[(P1)] $\E(Y_i)=y_i$ for every $i\in[n]$, and 
%     \item[(P2)] $\sum_{i\in[n]} Y_i \leq T$ with probability 1.
% \end{itemize}
% As a remark, the output of $\DR$ will also satisfy the negative correlation property of the dependent rounding developed by \citet{gandhi2006dependent}, although we do not make use of it explicitly.

\subsection{Approximation Algorithm \label{sec:ptkstatement}}

We propose the following algorithm for the ProbeTop-$k$ problem, which we call $\ALG_\ptk$. Given an instance, we first solve $\LP_{\ptk}(I)$. Let $x,y$ be an optimal solution. Define
\begin{equation}
    p_i = \frac{\sum_{j=1}^J x_{ij}}{y_i}, \quad v_i = \frac{\sum_{j=1}^J r_j x_{ij}}{\sum_{j=1}^J x_{ij}}. \label{eq:pandv}
\end{equation}
If any of the denominators are 0, then define these values as 0.\footnote{If $y_i=0$, then $\sum_{j=1}^J x_{ij}=0$ too. If $y_i=0$ then the algorithm will never interview applicant $i$, so the definition of these values is not relevant for the analysis.} For reasons that will soon become clear, $p_i$ is to be interpreted as the probability that we accept applicant $i$ given that they get interviewed, and $v_i$ is to be interpreted as the expected value of $i$ given that they get hired. Assume that after solving the LP, we relabel the applicants so that $v_1\geq v_2 \geq \cdots \geq v_n$. 

The algorithm first chooses which applicant to interview and the order in which the interviews are going to take place. For the first task, we use the dependent rounding DR just introduced. \orrev{Specifically, we feed vector $y$ as an input to DR, and obtain binary random variables $\{Y_i\}_{i\in [n]}$.} \orrev{The algorithm will choose to interview applicants $i$ with $Y_i=1$.  For the second task, the algorithm will always interview applicants in decreasing order of $v_i$.} 
%\sout{After deciding on the applicants to interview and the order, the interviews are conducted.}
Upon interviewing applicant $i$ and observing their value $r_j$, the algorithm hires $i$ with probability $x_{ij}/(y_i q_{ij})$. %A pseudo-code for this algorithm can be found in Appendix \ref{ap:pseudo}.

%As we are using DR to decide which candidates get interviews we have that if the optimal solution $y$ has zero or one fractional component, candidates are independently selected for receiving interviews. If there are two fractional components, then the two corresponding candidates have a perfect negative correlation (meaning that one is selected for interviewing if and only if the other one is not). We show that the algorithm performs better with this negative correlation than it would if each candidate was selected to be interviewed independently with probability $y_i$.\footnote{Selecting candidates this way could lead to unfeasible solutions where more than $T$ applicants receive interviews. Our algorithm with negative correlation outperforms the independent version even if it happens to violate this constraint.}

\subsection{\orrev{Analysis of $\ALG_\ptk$ }\label{sec:ptk_analysis}}

To analyze the algorithm let $Y_i$ be the indicator that applicant $i$ is included in the list of applicants to receive an interview. Let $Q_{ij}$ be the indicator of $V_i = r_j$. Let $P_i$ be the indicator that applicant $i$ would be hired if they were interviewed while still having unfilled positions. We colloquially refer to this event as $i$ `making the cut'. We have
\[\E(P_i) = \sum_{j=1}^J \E(P_i|Q_{ij})q_{ij} = \sum_{j=1}^J \frac{x_{ij}}{q_{ij} y_i}q_{ij} = p_i. \]

Define $Z_i = Y_i \cdot P_i$
as the indicator that $i$ is included in the list of applicants to receive an interview and {that $i$ makes the cut.}
%that the algorithm would respond with a hire decision upon revealing $i$'s value.
Note that $Y_i$ is independent of $P_i$. Let $N_{i-1} = \sum_{\ell=1}^{i-1} Z_{\ell}$ be the number of applicants that are interviewed before applicant $i$ that would be hired if there are still positions available. We can therefore write the (random) reward of our algorithm as $\sum_{i}  V_i \Ind\{N_{i-1} < k\}   Z_i $. The following lemma allows us to write the expected reward of $\ALG_\ptk$ in a way that will allow us to relate it to weighted rank functions of $k$-uniform matroids.
\begin{lemma}\label{lem:seq_exp_alg}
For any $I\in \I_{\ptk}$, we have $R_{\ALG_{\ptk}}(I) = \sum_{i=1}^n v_i \E(\Ind\{N_{i-1} < k\}Z_i)$.
\end{lemma}
{The proof of this lemma is deferred to \Cref{sec:seq_exp_alg_pf}.} Since the applicants are labeled such that $v_1\geq v_2 \geq \cdots \geq v_n$, \Cref{lem:seq_exp_alg} implies that our algorithm has the same expected reward as an algorithm that would first sample $Z_i = Y_i P_i$ for all applicants and then collect the $k$ highest values $v_i$ among those applicants with $Z_i=1$. This interpretation is only possible because $\ALG_\ptk$ is committed. The same expressions can be derived for an algorithm that instead of sampling $Y_i$ using DR, samples $\Tilde{Y}_i$ with the same marginal probabilities as $Y_i$, but independently. These expressions are useful for showing that our algorithm outperforms a \textit{hypothetical algorithm} that decides to interview each applicant $i$ using the independent indicators $\Tilde{Y}_i$ instead of the correlated indicators $Y_i$, potentially interviewing $T+1$ applicants.
The correlation induced by the constraint of $T$ interviews on the actual algorithm only works in our favor.

\begin{lemma}\label{lem:samplepath}
Let $y=(y_i)_{i\in[n]},x=(x_{ij})_{i\in[n],j\in[J]}$ be a basic feasible solution of $\LP_\ptk(I)$ with two fractional components $i_1$ and $i_2$. Let $\Tilde{Y}_i$ be independent Bernoulli random variables with mean $y_i$. Define $\Tilde{Z_i}=P_i \Tilde{Y}_i$ and $\Tilde{N}_i =  \min\{k,\sum_{\ell=1}^i \Tilde{Z}_\ell\}$. Then 
\begin{align}
    \sum_{i=1}^n v_i \E\left(Z_i \Ind\{ N_{i-1} < k \} \right)\geq \sum_{i=1}^n v_i \E\left(\Tilde{Z}_i \Ind\{ \Tilde{N}_{i-1} < k \}\right). \label{eq:indepineq}
\end{align}
\end{lemma}
{The proof of this lemma is deferred to \Cref{sec:samplepath_pf}.}

To conclude our analysis we relate the performance of the hypothetical algorithm (with independent indicators) and the objective function of the LP, through the weighted rank function of $k$-uniform matroids. For a set $S\subseteq[n]$ we define the weighted rank function for the $k$-uniform matroid with weights $v$ as  $v^*(S) = \max_{R\subseteq [S],\,|R|\leq k } \sum_{i\in R} v_i$.

Both the reward of the hypothetical independent algorithm and the objective of $\LP_\ptk$ can be expressed as expectations of weighted rank functions of $k$-uniform matroids. We use this together with the correlation gap results by \citet{yan2011mechanism} to show the main result of this section.
 
\begin{theorem}\label{thm:probetopk} For any instance $I\in \I_\ptk^k$, we have $R_{\ALG_\ptk}(I) \geq \left(1 - \frac{e^{-k}k^k}{k!}\right)\LP_{\ptk}(I)$.
\end{theorem}
{The proof of this theorem is deferred to \Cref{sec:probetopk_pf}.} In {\Cref{sec:tight}} we show that the guarantee that our algorithm attains is tight.

It is worth noting that this algorithm can be de-randomized. Indeed, the algorithm will randomize between at most two fixed orders, so both of them can be evaluated and the best among them can be selected. Thus, we have a deterministic, non-adaptive, committed algorithm whose expected reward will be at least a $(1 - e^{-k}k^k/k!)$ factor of the expected reward of the optimal algorithm. Since our algorithm is non-adaptive and committed, it establishes a lower bound on how good an approximation factor can be achieved by restricting to these classes of policies.

\subsection{\orrev{PTAS for ProbeTop-$k$} \label{sec:ptas}}

\orrev{We now explain how to combine $\ALG_\ptk$ with the approximation algorithm proposed by \citet{fu2018ptas} to obtain a PTAS for ProbeTop-$k$. We know there is a $(1-\varepsilon)$-approximation algorithm for ProbeTop-$k$ whose runtime is exponential in $k$ \citep[Theorem 4.2]{fu2018ptas}. The idea behind the combined PTAS is that for $k$ large enough, $\ALG_\ptk$ obtains an approximation factor better than $1-\varepsilon$, so for small values of $k$ we would run the algorithm by \citet{fu2018ptas}, and for large values of $k$ we would run our $\ALG_\ptk$. This essentially allows us to treat $k$ as a constant in the algorithm of \citet{fu2018ptas}.}
We note that our algorithm has runtime polynomial in $k$, but it is without loss to assume that $k$ is input in unary, since if $k$ (capacity) exceeds the number of given distributions (applicants/candidates) then the problem is trivial.

\orrev{Specifically, for $\varepsilon\in(0,1/e)$, let $k^*$ be the smallest $k\geq 1$ such that $\varepsilon>e^{-k}k^k/(k!)$.\footnote{\orrev{We know that $k^*$ always exists since $e^{-k}k^k/(k!)$ is a decreasing function that equals $1/e$ when $k=1$ and converges to 0 as $k\to\infty$.}} For $k<k^*$, the PTAS will run the algorithm by \citet{fu2018ptas} to obtain a $1-\varepsilon$ approximation in polynomial time, since $k^*$ is a constant. For $k\geq k^*$, the PTAS will run $\ALG_\ptk$ and obtain, in polynomial time, an approximation factor of $1-e^{-k}k^k/(k!) > 1-\varepsilon$.}
\begin{corollary}\label{cor:ptas}
\orrev{There is a PTAS for ProbeTop-$k$.}
\end{corollary}

\subsection{Weighted Bernoulli Values and the Sequential Offering Problem \label{subsec:weighted_ber}}

ProbeTop-$k$ is closely related to the Sequential Offering problem ($\seq$) studied by \citet{purohit2019hiring}. In this problem, the hiring firm has already interviewed $n$ candidates and must decide the order in which to send offers to them. Each candidate $i$ has a probability $q_i$ of accepting an offer and adds a value $r_i$ to the firm if they accept an offer. The firm can hire at most $k$ candidates and send at most $T$ offers in total. This model
% is \textit{almost}
is closely related to ProbeTop-$k$
% . Specifically, consider
when considering
the special case where the value of applicant $i$ takes value $r_i$ with probability $q_i$ (representing acceptance of an offer), and 0 with probability $1-q_i$ (representing rejection of an offer). The subtlety making $\ptk$ different from
$\seq$ is that policies for $\ptk$ can reject an applicant $i$ even if they had a realization to value $r_i$, which is not allowed in $\seq$.
% We show, however, that our algorithm can be made to not reject such candidates, without sacrificing performance.
We show, however, that $\ALG_\ptk$ can be {forced} to hire any applicant when their realized value is $r_i$ without loss, therefore making it an admissible algorithm for the Sequential Offering model with the same guarantee.

Indeed, we can rewrite $\LP_\ptk$ for this special case as
\begin{align*}
    \max \quad & \sum_{i=1}^n r_i x_{i} \nonumber\\
    s.t. \quad & x_{i} \leq y_i q_{i}\,\,\,\, \forall i \in [n], \quad \sum_{i=1}^n y_i \leq T, \quad \sum_{i=1}^n x_{i} \leq k, \quad x_{i}\geq 0 \,\,\,\, \forall i \in [n],  \quad  0 \leq y_i \leq 1 \,\,\,\, \forall i\in [n]
\end{align*}
% \begin{align*}
%     \max \quad & \sum_{i=1}^n r_i x_{i} \nonumber\\
%     s.t. \quad & x_{i} \leq y_i q_{i} & \forall i \in [n]  \\
%     & \sum_{i=1}^n y_i \leq T & \\
%     & \sum_{i=1}^n x_{i} \leq k& \\
%     & x_{i}\geq 0 & \forall i \in [n], \nonumber\\
%     & 0 \leq y_i \leq 1 & \forall i\in [n], \nonumber
% \end{align*}
where $x_i$ is to be interpreted as the probability of hiring applicant $i$ and $V_i=r_i$ (we can omit the variable corresponding to $V_i=0$ because it has a zero coefficient in the objective). The following lemma will let us restrict $\ALG_\ptk$ to be admissible for the Sequential Offering problem.
\begin{lemma} \label{lem:equivalent_sol_for_seq}
For weighted Bernoulli instances, $\LP_\ptk$ has an optimal solution with $x_i = y_iq_i \ \forall i\in[n]$.
\end{lemma}
{The proof of this lemma is deferred to \Cref{sec:equivalent_sol_for_seq_pf}.}

Recall that after interviewing $i$, $\ALG_\ptk$ will hire them with probability $x_i/(y_i q_i)$. Therefore, if we restrict to solutions with $x_i = y_iq_i$, then $\ALG_\ptk$ will hire applicant $i$ with probability 1 if $V_i =r_i$, making it admissible for the Sequential Offering problem.

{For this specific family of instances, equations (\ref{eq:pandv})} yield
\[ p_i = \frac{x_i}{y_i} = x_i \frac{q_i}{x_i} = q_i, \quad v_i = \frac{r_i x_i}{x_i} = r_i. \]
We can {use these expressions to} further simplify the LP by removing variables $x_i$ and express it in terms of $p_i$ and $v_i$, leading to $\LP_\seq$.
\begin{align*}
    \LP_\seq(I) = \max \quad & \sum_{i=1}^n v_i y_{i} p_i \nonumber\\
    s.t. \quad &  \sum_{i=1}^n y_i \leq T, \quad \sum_{i=1}^n y_{i}p_i \leq k, \quad 0 \leq y_i \leq 1 \,\,\,\, \forall i\in [n].
    % & \sum_{i=1}^n y_{i}p_i \leq k& \\
    % & 0 \leq y_i \leq 1 & \forall i\in [n]. \nonumber
\end{align*}

Being consistent with the previously introduced notation, let $\I_\seq$ be the set of all instances for the Sequential Offering problem. Let $\I_\seq^k$ be the set of all instances of $\seq$ that have exactly $k$ positions to fill. Let $\ALG_\seq$ be the modified version of $\ALG_\ptk$ that, when faced with instances of weighted Bernoulli random variables, modifies the solution of the LP {as in \Cref{lem:equivalent_sol_for_seq}} such that the algorithm is admissible for $\seq$. We then have the following corollary of \Cref{thm:probetopk}.

\begin{corollary}\label{cor:seq} For any instance $I\in \I_\seq^k$, we have $R_{\ALG_\seq}(I) \geq \left(1 - \frac{e^{-k}k^k}{k!}\right)\LP_{\seq}(I)$.
\end{corollary}

\section{Parallel Offering \label{sec:paral}}

We turn to the Parallel Offering model defined in \Cref{sec:parintro}. We provide an algorithm that achieves a $(1-1/e)$ approximation of the optimal policy. The algorithm works by solving a LP relaxation and rounding its solution to decide who to offer which position, and in which order.

In \Cref{sec:parLP} we introduce the LP in question. In \Cref{sec:parALG} we present {our algorithm and in \Cref{sec:par_analysis} we provide its analysis.} In \Cref{sec:batching} we study the special case with identical positions introduced by \citet{purohit2019hiring} and establish a connection between the Parallel and Sequential Offering problems.
% The proofs and complementary materials from this section can be found in \Cref{ap:paral}.

\subsection{Linear Program \label{sec:parLP}}

For an instance $I\in\I_\paral$, we introduce $\LP_\paral(I)$, with LP variables $y_{ij}$ for $i\in[n]$ and $j\in[k]$. Variable $y_{ij}$ is to be interpreted as the probability that candidate $i$ receives an offer for position $j$. As with $\LP_\ptk$, this LP only enforces that the problem's constraints are satisfied in expectation.
\begin{align}
    \LP_\paral(I) = \max \quad &  \sum_{j=1}^k \sum_{i=1}^n v_{ij} y_{ij}  p_{ij} \nonumber\\
    s.t. \quad & \sum_{i=1}^n y_{ij} \leq T & \forall j\in[k] \label{const:time_limit}\\
     & \sum_{i=1}^n y_{ij}p_{ij} \leq 1 & \forall j\in[k] \label{const:acceptance_limit}\\
    & \sum_{j=1}^k y_{ij} \leq 1 & \forall i\in[n]\label{const:offer_limit}\\
    & 0\leq  y_{ij} \leq 1 & \forall (i,j)\in[n]\times [k].\nonumber
\end{align}
We start by formally proving that $\LP_\paral$ upper-bounds the expected reward of any algorithm.
\begin{lemma} \label{lem:paral_ub}
For any instance $I\in\I_\paral$, we have $\LP_\paral(I)\geq \OPT_\paral(I)$.
\end{lemma}
{The proof of this lemma is deferred to \Cref{sec:paral_ub_pf}.}

\subsection{Approximation Algorithm \label{sec:parALG}}

We now present our algorithm for the Parallel Offering model, which we refer to as $\ALG_\paral$. The algorithm first solves $\LP_\paral$ to produce an optimal solution $y$. With the solution at hand, the algorithm will round it to obtain a random binary matrix $Y = (Y_{ij})_{(i,j)\in [n]\times[k]}$.
To round our solution here, we use the dependent rounding scheme developed by \citet{gandhi2006dependent} that is described in \Cref{app:gkps_rounding}. \orrev{Specifically, the nodes of one side of the bipartite graph are the $n$ applicants, and the nodes of the other side are the $k$ positions. The weight of each edge $(i,j)\in[n]\times [k]$ is the fractional value of $y_{ij}$ from the optimal solution of $\LP_\paral$.} After the solution is rounded, the algorithm uses the rounded solution to make a sequential offering list for each position so that candidate $i$ is included in the list for position $j$ if $Y_{ij}=1$. The properties of the {dependent} rounding scheme by \citet{gandhi2006dependent} combined with the constraints of $\LP_\paral$ will ensure that: (a) each list has at most $T$ candidates, and (b) each candidate is included in at most one list. After the lists are formed, {all lists are run} in parallel, and the order in which candidates $i$ for each position $j$ are sent offers is in decreasing order of $v_{ij}$. %A pseudo-code for this algorithm can be found in Appendix \ref{ap:pseudo}.

\subsection{\orrev{Analysis of $\ALG_\paral$ }\label{sec:par_analysis}}

For the analysis, let $L_j$ denote \orrev{the expected value of the candidate that ends up being hired for position $j$.}
%We will show that for each position $j$ it holds that $L_j\geq (1-1/e)L^*_j$.
Define $Z_{ij} = P_{ij} Y_{ij}$ and $\Tilde{Z}_{ij} = P_{ij} \Tilde{Y}_{ij}$, where \orrev{$(\Tilde{Y}_{ij})_{i\in[n],j\in[k]}$} are independent Bernoulli random variables with  $\E(\Tilde{Y}_{ij}) = y_{ij}$. Let $D_{j} = \{ i : Z_{ij} =1\} $ and $\Tilde{D}_j =\{ i : \Tilde{Z}_{ij} =1\} $. Let $v_j^*(S) = \max_{i\in S} v_{ij}$. It is clear to see that $L_j = \E(v_j^*(D))$.

The first step for showing the guarantee of $\ALG_\paral$ is to show that the reward collected by a list is not lower than what we would collect if we rounded each component of $y$ independently. This is a consequence of the negative correlation property (P3) of the dependent rounding scheme by \citet{gandhi2006dependent}. This result is formalized in the following lemma.

\begin{lemma}\label{lem:paral_indep} For all $j\in[k]$, we have $\E(v_j^*(D))\geq\E(v^*_j(\Tilde{D}))$.
\end{lemma}
{The proof of this lemma is deferred to \Cref{sec:paral_indep_pf}.}
\orrev{
We note that \Cref{lem:paral_indep} analyzes each position in isolation, even if two or more positions are identical.
In fact, our proof does not appear to extend to multiple identical positions.
Indeed, our proof only makes use of properties (P1) and (P3) from the dependent rounding scheme of \citet{gandhi2006dependent}.
We prove in \Cref{app:counterexample} that only using these properties is insufficient for proving an analogue of \Cref{lem:paral_indep} for multiple positions.
}

% In the proof of \Cref{lem:paral_indep} we only use (P1) and (P3) from the rounding scheme by \citet{gandhi2006dependent}. These properties are not enough if we want to show that the dependent rounding outperforms the independent rounding when selecting the $k\geq 2$ elements with the highest weights (which holds, for instance, in \Cref{lem:samplepath}). See \Cref{eg:negCorrelWorseThanIndep} in \Cref{app:counterexample} for a counterexample.

Continuing with the main result, \orrev{let $L^*_j$ denote $\sum_{i=1}^n v_{ij} y_{ij} p_{ij} $, so that the objective function of $\LP_\paral$ can be expressed as $\sum_{j=1}^k L^*_j$.} The following lemma helps establish the desired bound for each {separate} list; note that at most one candidate is hired from each list. Both the correlation gap results and the rounding scheme mentioned in \Cref{sec:known_results} are used in our analysis. 
\begin{lemma}\label{lem:list_bound}
For all $j\in[k]$, we have $\E(v^*_j(\Tilde{D})) \geq (1-1/e) L_j^*$.
\end{lemma}
{The proof of this lemma is deferred to \Cref{sec:list_bound_pf}.}

By combining \Cref{lem:paral_indep} and \Cref{lem:list_bound} we obtain the main result of the section.
\begin{theorem}\label{thm:paral}
For any instance $I\in\I_\paral$, we have $R_{\ALG_\paral}(I)\geq(1-1/e)\LP_\paral(I)$.
\end{theorem}
In \Cref{sec:parTight} we establish that the bound achieved by our algorithm is tight.

\subsection{Identical Positions and the Cost of Batching \label{sec:batching}}

We turn to the special case where all positions are identical (i.e. $v_{ij} = v_i$ and $p_{ij} = p_i$ for all $i\in[n]$ and $j\in[k]$). For this case, we can obtain a connection between the Sequential Offering model and the Parallel Offering model through their respective LPs.

 \orrev{Let $\I_\paral^=\subseteq \I_\paral$ be the set of instances of $\paral$ in which all positions are identical. For a given instance $I\in\I_\paral^=$, construct $I'\in\I_\seq$ \orrev{with the same candidates} as \orrev{in} $I$, but with a budget of $kT$ sequential offers (instead of $T$ parallel offering rounds)}. We obtain the following lemma.
\begin{lemma}\label{thm:cost_of_batching}
$\LP_\paral(I) = \LP_\seq(I')$.
\end{lemma}
{The proof of this lemma is deferred to \Cref{sec:cost_of_batching_pf}.}

\Cref{thm:cost_of_batching} implies the following corollary, which we refer to as the cost of batching.
\begin{corollary}\label{cor:cost_of_batching}
For any \orrev{$I\in \I_\paral^=$}, we have $\ALG_\paral(I) \geq (1-1/e)\OPT_\seq(I')$.
\end{corollary}
This corollary helps us understand how costly it can be to send offers in batches instead of one by one like in the Sequential Offering problem. By reducing to $T$ parallel offering rounds instead of $kT$ sequential offering rounds, we know that we cannot be worse by more than a factor of $(1-1/e)$.

\section{Simultaneous Offering \label{sec:simul}}
%\linespread{\linespace}\selectfont{}

In this section, we study the Simultaneous Offering problem described in \Cref{sec:simul_intro}. We are interested in a class of "value-ordered" policies.  \orrev{We develop such a policy that, when faced with instances where all values of candidates are lower-bounded by some $\tau\in(0,1)$, achieves an $\alpha_k^\tau$-approximation, where $\alpha_k^\tau$ is an increasing function that maps the number of positions $k\in \mathbb{N}$ and the lower bound $\tau$ to a real number between 0 and 1}. Our algorithm first solves an LP relaxation.  It then uses a modification of the optimal solution to decide which candidates will receive offers.

\orrev{In \Cref{sim:policies} we define the class of value-ordered policies, which we show can perform arbitrarily poorly without the assumption of the lower bound $\tau$. In \Cref{sec:simLP} we introduce an LP relaxation that is used in our algorithm. {In \Cref{sec:sim_alg,sec:simLB} we introduce and analyze our algorithm,}
\orrev{providing lower bounds on how well value-ordered policies can perform.}}
% and characterizing the region in which these bounds coincide.}
%introduce our value-ordered policy and its lower bound, provide an upper bound for how well value-ordered policies can perform, and characterize the region where these two bounds coincide.
%The proofs and complementary materials of this section are located in \Cref{ap:simul}.

\subsection{Value-ordered Policies \label{sim:policies}}
%\linespread{\linespace}\selectfont{}

For this problem, we focus on a natural class of policies, which we call \textit{value-ordered policies}. Assume that $v_1 \geq v_2 \geq \cdots \geq v_n$. A value-ordered policy is defined by an integer $m\in[n]$, and sends offers to the $m$ candidates with the highest values (i.e. $S=\{1,\dots,m\}$). \orrev{This family of policies is practically well-motivated because a firm generally does not want to withhold sending offers to high-value candidates whom it deems "too good" for itself. The optimal value-ordered policy also can be obtained efficiently \orrev{by solving a dynamic program (see \Cref{app:simul_DP} for details)}.}

%\orrev{Indeed, it would be unnatural in practice for a firm to not send an offer to a candidate that they perceive as better than other candidates that do receive offers, just because they think it is too unlikely that they accept. }%as firms ideally want to make offers to the $m$ highest-value candidates without considering their probabilities of acceptance.

As natural as they seem, value-ordered policies can achieve an arbitrarily poor approximation factor. Consider the following simple example.
\begin{example} \label{eg:simul_VO}
Consider an instance with $k=1$ and $n=2$. For small $\varepsilon$, candidate 1 has $v_1 = p_1 = \varepsilon$, while candidate 2 has $v_2 = \varepsilon(1-\varepsilon)$ and $p_2 =1$. There are three possible policies for this instance: $\{1,2\}$, $\{1\}$ and $\{2\}$. The two first are value-ordered policies. The reward for using $\{1,2\}$ is 0, for using $\{1\}$ is $\varepsilon^2$, and for using $\{2\}$ is $\varepsilon(1-\varepsilon)$. For $\varepsilon<1/2$ this means that the optimal policy is $\{2\}$ and the optimal value-ordered policy is $\{1\}$.
The approximation factor achieved by value-ordered policies in this instance is equal to $\varepsilon/(1-\varepsilon)$, which can be made arbitrarily small by taking $\varepsilon \to 0$.

% First notice that $p_2=1$ and $k=1$, so sending an offer to both candidates will never be optimal. Since $\{1,2\}$ and $\{1\}$ are the only two value-ordered policies, this means that $\{1\}$ is the optimal value-ordered policy. \wnote{I'm not sure I follow the logic here.  I agree that you would never offer 1 if you already offered 2, but an optimal \textit{value-ordered} policy may still want to offer both?} The expected reward for sending an offer only to candidate 1 is $\varepsilon^2$. On the other hand, the expected reward for only sending an offer to candidate 2 is $\varepsilon(1-\varepsilon)>\varepsilon^2$ for $\varepsilon<1/2$, so the optimal policy is to only send an offer to candidate 2. 
\end{example}
A similar example can be constructed given any amount of positions $k$, as we {show later in \Cref{thm:simul_upper_bound}}. The intuition behind the previous example is that when $\varepsilon$ is small, adding a candidate over capacity provides negligible benefit compared to cost, making the capacity $k$ almost a hard constraint. 
\orrev{And although the difficulties of this specific example can be avoided by ordering the candidates in descending order of $v_ip_i$ instead of $v_i$, or using the greedy policy which starts with an empty set of candidates and iteratively adds the candidate with the highest marginal benefit to the set, we show that these alternate algorithmic ideas can
also achieve an arbitrarily poor approximation in \Cref{app:EO_example}.}

% Since this last class of policies can perform arbitrarily bad and they are less practically motivated, we will restrict our attention to value-ordered policies.

We hereafter aim for constant factor guarantees for value-ordered policies under the assumption that the values of the candidates are lower-bounded by a parameter $\tau > 0$. Let $\I_\simul^\tau =\{ I \in \I_\simul : v_i \geq \tau \,\, \forall i\in [n] \}$ be the set of $\tau$-bounded instances. \orrev{Since we normalized $c$ (the cost of hiring each candidate over capacity) to be 1, the value of $\tau$ can be interpreted as how hard or soft the capacity constraint of $k$ is in practice. For example, if $\tau$ is close to 1, then the capacity constraint can be viewed as soft, since the cost of hiring over capacity would be almost fully compensated by the value of any candidate.
If the capacity constraint is hard in practice, then $\tau$ could be close to 0.}

\orrev{
We remark that if $\tau$ is only small because of "irrelevant" low-value candidates who would never receive an offer, then it does not negatively affect our results.  More precisely, our guarantee depends on the lowest value of a candidate with positive mass in the solution of the LP introduced in \Cref{sec:simLP}.  We ignore this distinction in our definition of $\tau$-bounded instances for simplicity.
} 
% In contrast, if $\tau$ is close to 0, then the capacity constraint is rather hard.  That being said, a low value of $\tau$ does \footnote{\orrev{A low value of $\tau$ does not necessarily imply that the capacity constraint is almost hard. This might not be the case in an instance where there is a single low-value candidate (thus low $\tau$) who we will never consider because there exist enough candidates with a value close to 1 so that we deem the low-value candidate as irrelevant.}}}

% then there exist candidates who will never be worth the risk of hiring over capacity, and the capacity constraint resembles more of a hard constraint.}

We say that a policy $\pi$ is an $\alpha$-approximation for $\tau$-bounded instances if 
\[\inf_{I \in \I_\simul^\tau} \frac{R_\pi(I)}{\OPT_\simul(I)}\geq \alpha.\] 
Notice that if $\tau\geq 1$ then the problem is trivial: it is optimal to send an offer to every candidate. We will therefore restrict our attention to $\tau \in (0,1)$.

Before we carry on, we define $V_\high = \{i\in[n]: v_i > 1\}$ to be the set of candidates whose value is greater than 1. These are candidates that we would like to hire even if we know they would violate capacity{, essentially making the number of positions random, as they will all receive offers but it is uncertain how many would accept. To our understanding, there is no easy way to reduce the problem to one where they do not exist, when both $V_\high$ and $[n]\setminus V_\high$ are non-empty.}

%We believe it is important to model such candidates, and   
% {These kind of candidates do appear in practice and there is no obvious way of reducing the problem to a setting where $V_\high$ is empty.}

\subsection{LP Relaxation \label{sec:simLP}}

In order to obtain approximation factors for value-ordered policies, we introduce a linear programming relaxation of $\OPT_\simul$ which we call $\LP_\simul$. 
\begin{align}
    \LP_\simul(I) = \max_{y,z} \quad& \sum_{i\in [n]} v_ip_iy_i + z \nonumber\\
    s.t. \quad & z \leq k - \sum_{i\in[n]} p_i y_i \nonumber\\
    & z \leq 0\nonumber\\ 
    &0\leq  y_i \leq 1 \,\,\,\, \forall i\in[n]\nonumber.
\end{align}
% \begin{align}
%     \LP_\simul(I) = \max_{y,z} \quad& \sum_{i\in [n]} v_ip_iy_i + z \nonumber\\
%     s.t. \quad & z \leq k - \sum_{i\in[n]} p_i y_i \nonumber\\
%      &  z \leq 0 \nonumber \\
%     & 0\leq  y_i \leq 1 & \forall i\in[n].\nonumber
% \end{align}
\orrev{As with the previous linear programs, $y_i$ is to be interpreted as the probability that candidate $i$ receives an offer. This linear program optimizes over randomized policies that pay a penalty for the difference between the expected number of candidates hired and $k$, rather than the realized number of candidates hired in excess of $k$.} Thus, by applying Jensen's inequality we can show that $\LP_\simul$ is an upper bound of $\OPT_\simul$.
\begin{lemma}\label{lem:sim_LP_ub}
For any $I \in \I_\simul$, we have $\LP_\simul(I) \geq \OPT_\simul(I)$.
\end{lemma}
{The proof of this lemma is deferred to \Cref{sec:sim_LP_ub_pf}.}

We proceed to show a property about optimal solutions of $\LP_\simul$ that closely relates it to a value-ordered policy.

\begin{lemma}\label{lem:simDecreasingSol}
Let $I\in \I_\simul$. There exists an optimal solution $(y,z)$ of $\LP_\simul(I)$ with an index $j$ such that $y_i =1$ for $1\leq i \leq j-1$, $y_j >0$, and $y_i =0$ for $j+1 \leq i \leq n$.
\end{lemma}
\orrev{The proof of this lemma follows from the fact that, for any fixed $z$, $\LP_\simul$ corresponds to an instance of the Fractional Knapsack problem. It is well-known that optimal solutions to this problem satisfy the structure described in \Cref{lem:simDecreasingSol} \citep[Chapter 5]{goodrich2001algorithm}, so it is true for any optimal choice of $z$.}
%\orrev{The proof of this lemma is deferred to \Cref{sec:simDecreasingSol_pf}.}
\orrev{Given \Cref{lem:simDecreasingSol}, we can easily construct a randomized value-ordered policy from an optimal solution of $\LP_\simul$.}
%This lemma allows us to use an optimal solution of $\LP_\simul$ to construct a randomized value-ordered policy.
\orrev{Indeed, we could {simply} send an offer to each candidate $i$ independently with probability $y_i$. This policy randomizes between two value-ordered policies: sending offers to candidates $\{1,\dots,j\}$ or to candidates $\{1,\dots,j-1\}$. {We will make use of this idea when constructing our actual approximation algorithm.}}
% Given the structure revealed in Lemma \ref{lem:simDecreasingSol}, this is the same as sending offers to candidates $\{1,\dots,j\}$ with probability $y_j$, and to candidates $\{1,\dots,j-1\}$ with probability $1-y_j$. \wnote{This previous sentence is not the correct description of the policy? I think so}  This policy randomizes between two value-ordered policies, 
%Therefore, a lower bound on its performance immediately implies a lower bound on the performance of the optimal value-ordered policy.
%In the following subsection, we introduce $\ALG_\simul^s$, a policy that follows the same idea as the policy just described, but allows a parameter $s\in[0,1]$ to further tune and improve its performance. 

\orrev{An important object in the definition and analyses of our policies is the ``total mass'' of a solution of $\LP_\simul$.} For a feasible solution $y$, {its total mass is given by} $\sum_{i\in [n]}y_i p_i$. This can be interpreted as the expected amount of candidates that would be hired if we were to send an offer to each candidate $i$ with probability $y_i$. \Cref{lem:simDecreasingSol} also implies that an optimal solution is completely determined by its total mass. This is because the solution can be constructed by ``filling'' the components from smallest to largest index {until we obtain the desired total mass}.

%The following lemma also sheds light on the structure of optimal solutions of $\LP_\simul$ by identifying the total mass of an optimal solution depending on the instance parameters. Since the total mass completely determines an optimal solution, this result is useful for computing optimal solutions and also for establishing the guarantees of $\ALG_\simul^s$.

\orrev{The following lemma allows us to determine the total mass of optimal solutions of $\LP_\simul$ based on aggregate parameters of instances and will be useful later in the analysis.}
\begin{lemma}\label{lem:simul_mass_solution}
Let $I\in \I_\simul$. Let $(y,z)$ be an optimal solution of $\LP_\simul(I)$ that satisfies the structure given in \Cref{lem:simDecreasingSol}. Let $P_\high = \sum_{i\in V_\high} p_i$ and $P_\tot = \sum_{i \in [n]} p_i$. The following holds:

\begin{enumerate}
    \item If $P_\high>k$, then $y_i = 1$ for all $i \in V_\high$ and $y_i=0$ otherwise. Therefore, $\sum_{i\in [n]} y_i p_i = P_\high$;
    \item If $P_\high \leq k$, then $\sum_{i\in [n]} y_i p_i = \min\{k,P_\tot\}$.
\end{enumerate}
\end{lemma}
{The proof of this lemma is deferred to \Cref{sec:simul_mass_solution_pf}.}

%This lemma gives us a simple way of computing an optimal solution for $\LP_\simul$. If $P_\high>k$, then set $y_i=1$ for all $\in V_\high$, and 0 otherwise. If $P_\high \leq k$, then we can simply ``fill'' the solution from smallest to largest index, until either $\sum_{i\in[n]} y_i p_i =k$ or $y_i=1$ for all $i\in [n]$. In either case, we set $z=\min\{0, k - \sum_{i\in[n]}y_i p_i\}$. This way, there is no need to solve the LP when implementing $\ALG_\simul^s$.

\subsection{Approximation Algorithm\label{sec:sim_alg}}
% \linespread{\linespace}\selectfont{}
Let us introduce our approximation algorithm for $\simul$, which we call $\ALG_\simul^s$. The policy takes as input an optimal solution $y$ of $\LP_\simul$ and a parameter $s\in[0,1]$. The idea of the policy is to truncate the optimal solution of the LP. This truncation is done by scaling down the total mass of the optimal LP solution by a factor $s$ and then using this mass to ``fill'' the new variables from smallest to largest index. We then use this truncated solution to decide which candidates will receive offers. 

Formally, the first step is to construct an alternative solution $y'$ by truncating $y$ the following way. Let $\kappa = s\sum_{i\in [n]} y_i p_i$ be the total mass of the optimal LP solution, scaled down by a factor $s$. Let $j$ be the first index such that $ \sum_{i=1}^{j+1} p_i > \kappa$. We let $y_i'=1$ for $1\leq i \leq j$, $y_i'=0$ for $j+2 \leq i \leq n$, and set $y_{j+1}'$ such that $\sum_{i=1}^{j+1} y_i' p_i = \kappa$. After solution $y'$ is constructed, the offers are sent independently to each candidate $i$ with probability $y_i'$. Given the structure of $y'$, $\ALG_\simul^s$ clearly randomizes between two value-ordered policies: sending offers to candidates 1 through $j$, and sending offers to candidates 1 through $j+1$. %A pseudo-code for this algorithm can be found in Appendix \ref{ap:pseudo}. 
{For any $k$ and $\tau$, parameter $s$ can be optimized to maximize the algorithm's performance, and we let $s^*$ denote the optimal value.}
%\orrev{For given $\tau$ and $k$, we will be interested in setting $s$ in order to optimize the algorithm's performance, and we will call this optimal parameter choice $s^*$. }

% In order to analyze $\ALG_\simul$ we recognize three different cases, which lead to three different solutions for $\LP_\simul$. The three cases are when 

\subsection{\orrev{Analysis of $\ALG_\simul$} \label{sec:simLB}}

\orrev{The analysis of $\ALG_\simul^s$ \orrev{and $\ALG_\simul^{s^*}$} is done in two cases. One case is $\sum_{i\in V_\high} p_i \leq k$, in which sending an offer to all candidates in $V_\high$ hires an expected number of candidates less than or equal to $k$. The other case is $\sum_{i\in V_\high} p_i > k$.}  %We start by analyzing the latter and then show that the former case is not harder than the latter for an optimally chosen parameter $s$.

\orrev{\textbf{Case 1: $\sum_{i\in V_\high} p_i \leq k$.}} In this case, we can show a lower bound for the performance of $\ALG_\simul^s$ that {depends on the} choice of $s$.
\begin{lemma} \label{lem:simul_fixed_s}
For any $s\in[0,1]$, $\tau\in(0,1)$, and $I\in \I_\simul^\tau$ with $\sum_{i\in V_{\high}} p_i \leq k$, we have
\[ R_{\ALG_\simul^s}(I) \geq \left( s - \frac{s}{\tau} + \frac{\E[\min\{\poiss(sk),k\}]}{\tau k}  \right)\LP_\simul(I). \]
\end{lemma}
{The proof of this lemma is deferred to \Cref{sec:simul_fixed_s_pf}.}

% The previous theorem implies that setting $s=1$ gives a lower bound of $1 - e^{-k}k^k/(k!\tau)$. We can obtain the optimal parameter $s^*$ that achieves the best possible lower bound.

% Let $\ALG_\simul^* = \ALG_\simul^{s^*}$. \wnote{Do we need to define this?  It seems like you don't even always use it, e.g. in the statement of Lemma 13}

\orrev{\textbf{Case 2: $\sum_{i\in V_\high} p_i > k$.}} In this case, we can show that for an optimally chosen parameter $s^*$, the guarantee obtained by $\ALG^{s^*}_\simul$ cannot be worse than the previous case where $\sum_{i\in V_{\high}} p_i \leq k$. A key observation for analyzing this new case is that as \Cref{lem:simul_mass_solution} states, all candidates $i$ whose value $y_i$ in the optimal LP solution will have $v_i\geq 1$, so $s^*=1$. Indeed, setting $s<1$ can only decrease the chances of sending offers to candidates in $V_\high$, and these are candidates that the firm would always want to hire (even over capacity). With this observation in hand,
we prove the following, via the construction of an auxiliary instance in which the probabilities of acceptance are deflated.
% we can construct an auxiliary instance by maintaining the values, but deflating the probabilities of acceptance such that we go back to the case of Lemma \ref{lem:simul_fixed_s}.

% Given the structure in Lemma $\ref{lem:simul_mass_solution}$, the optimal solution will be to set $y_i=1$ for all $i \in V_\high$ and $y_i=0$ for the remaining candidates. Intuitively, all candidates with positive solution will have $v_i >1$, so even if we hire over capacity, the marginal value of hiring any of these candidates will be positive. This discussion is formalized in the proof of the Lemma.

\begin{lemma}\label{lem:simul_easy_case}
Let $I = (k,n,p,v) \in \I_\simul^\tau$ be an instance with $\sum_{i\in V_\high} p_i >k$. Then there exists an instance $I' = (k,n,p',v) \in \I_\simul^\tau$ such that $\sum_{i\in V_\high} p_i' = k$  and
\[ \frac{R_{\ALG_\simul^{s^*}}(I)}{\LP_\simul(I)} \geq \frac{R_{\ALG_\simul^{s^*}}(I')}{\LP_\simul(I')} \]

\end{lemma}
{The proof of this lemma is deferred to \Cref{sec:simul_easy_case_pf}.}

By combining \Cref{lem:simul_fixed_s} and \Cref{lem:simul_easy_case} for an optimally chosen parameter $s^*$, we obtain the main result of this section.

\begin{theorem}\label{thm:simul_lower_bound}
For any $\tau\in(0,1)$ and $I \in \I_\simul^\tau$ we have $R_{\ALG_\simul^{s^*}}(I) \geq  \alpha_k^\tau \LP_\simul(I)$, where
\[ \alpha_k^\tau = \sup_{0\leq s \leq 1} \left( s - \frac{s}{\tau} + \frac{\E[\min\{\poiss(sk),k\}]}{\tau k}  \right). \]
\end{theorem}

\orrev{The performance of $\ALG_\simul^{s^*}$ clearly dominates that of $\ALG_\simul^1$, where $\ALG_\simul^1$ directly follows the LP solution. By \Cref{lem:simul_fixed_s}, $\ALG_\simul^1$ has a guarantee of at least $1 - \frac{1}{\tau} + \frac{\E[\min\{\poiss(k),k\}]}{\tau k}$, which equals $1 - e^{-k}k^k/(k!\tau)$, as we derive in \orrev{\Cref{sec:TruncPois}}.} \orrev{By combining this last expression with Stirling's approximation, we obtain the following result about the asymptotic optimality of $\ALG^{s^*}_\simul$ when the number of positions grows large and the values of candidates are bounded away from 0.}
\begin{corollary}
    \orrev{For any $\tau\in(0,1)$ and any instance $I\in\I^\tau_\simul$, we have
    $ \ALG_\simul^{s^*} \geq (1-O(1/\sqrt{k}))\LP_\simul(I)$, where $k$ is the number of positions in the instance.} \label{thm:simul_asymptotic}
\end{corollary}

% A remark about our lower bounds is that they hold in a stronger setting. Indeed, all of our algorithms and analyses only consider candidates $i$ whose value of $y_i$ in the optimal LP Solution is positive.
% Define $\I_\simul^{\tau, \LP}$ as the set of instances for which all of the candidates with positive LP solution have value bounded from below by $\tau$. All of our results hold if we change $\I_\simul^\tau$ by $\I_\simul^{\tau, \LP}$, and clearly $\I_\simul^{\tau} \subseteq \I_\simul^{\tau, \LP}$. This illustrates that the guarantee of our algorithms is not damaged by low-value candidates who are irrelevant because they are never going to be considered.

We now provide an upper bound for the guarantee that can be achieved using value-ordered policies. We construct an instance consisting of $n = j^2 + k$ candidates\footnote{We \orrev{do not} need exactly $j^2$ candidates of type 1 for showing the result. Any amount of candidates $\ell_j$ with $\ell_j/j \to \infty$ as $j\to\infty$ will suffice.}, with $j$ being a (large) integer. The first $j^2$ candidates are of type 1, who have $p_i = k/j$ and $v_i= \tau + \delta$, where $\delta >0$ is small. The remaining $k$ candidates are of type 2, who have $p_i=1$ and $v_i = \tau$. The idea behind this result is that any optimal value-ordered policy will only send offers to type 1 candidates. On the other hand, the optimal policy does not perform worse than a policy that sends offers to $k$ type 2 candidates and zero type 1 candidates.

\begin{theorem}\label{thm:simul_upper_bound}
For any $\varepsilon>0$ and $\tau \in (0,1)$, there exists an instance $I \in \I_\simul^\tau$ such that no value-ordered policy can have an expected reward greater than
$\left( \beta_k^\tau + \varepsilon  \right)\OPT_\simul(I)$, where
\[ \beta_k^\tau = \sup_{s\geq 0}\left( s - \frac{s}{\tau} + \frac{\E[\min\{\poiss(sk),k\}]}{\tau k} \right) . \]
\end{theorem}
{The proof of this theorem is deferred to \Cref{sec:simul_upper_bound_pf}.}  \orrev{In \Cref{app:simil:plots} we analyze the region in which our bounds coincide, together with plots showing $\alpha_k^\tau$ and $\beta_k^\tau$ for several values of $k$ and $\tau$.}

\section{Future Directions \label{sec:donclusion}}
%\linespread{\linespace}\selectfont{}

%In this paper, we study several problems motivated by a firm's hiring pipelines. For sequential interviewing, sequential offering and parallel offering we develop LP-based policies that improve the state of the art in both approximation guarantees, adaptivity gaps and costs of batching. We also introduce the Simultaneous Offering problem and characterize the performance of practically motivated value-ordered policies. We provide numerical experiments that give insights on the factors a firm should take into consideration when deciding which offering procedure to adopt.

\orrev{We close the paper by pointing out natural follow-up research questions that arise from our work.} An interesting open direction is to combine the sequential interviewing problem with the offering problem. This can {be done} by generalizing the sequential interviewing problem to a setting where candidates are not assured to accept offers and instead have a probability of accepting which might depend on factors such as their realized value. The firm can, at the end of each period, decide to send offers to any candidate who has already been interviewed. Another open direction is to combine the Simultaneous Offering problem with the Parallel Offering problem, as implicitly suggested in \citet{purohit2019hiring}. In this setting, the firm could, at each round, send more offers than positions remaining and face the risk of hiring over capacity (at a cost). One last future direction on the modeling side is how the firm should behave when the problem parameters are inaccurate. In particular, acceptance probabilities can be very difficult to estimate, so developing algorithms that are robust to these parameters' misspecification could be of great interest to firms.

\orrev{On the technical side, we believe the Simultaneous Offering problem introduced is a parsimonious new variant of overbooking.  Although our focus was to analyze the performance of value-ordered policies, we do not know of any hardness or algorithmic results for finding the best offer set.  It would be interesting if an optimal or near-optimal (i.e., PTAS) algorithm could be found.}

\subsubsection*{Acknowledgements.}
The authors thank
Jos\'e Correa for insightful discussions about Sequential Offering,
Rad Niazadeh for pointing us to the highly relevant reference \citet{bradac2019near},
and Aravind Srinivasan for insightful discussions about negative association.
The authors further thank anonymous reviewers for \textit{Operations Research} who gave exceptionally detailed comments and identified the corollaries in \Cref{sec:ptas,sec:simLB}.

\SingleSpacedXI
% \DoubleSpacedXI

% Acknowledgments here
%\ACKNOWLEDGMENT{The authors gratefully acknowledge the existence of
%the Journal of Irreproducible Results and the support of the Society
%for the Preservation of Inane Research.}

\bibliographystyle{informs2014} % outcomment this and next line in Case 1
% \linespread{\linespace}\selectfont{}
\bibliography{main} % if more than one, comma separated

\begin{thebibliography}{26}
\providecommand{\natexlab}[1]{#1}
\providecommand{\url}[1]{\texttt{#1}}
\providecommand{\urlprefix}{URL }

\bibitem[{Agrawal et~al.(2010)Agrawal, Ding, Saberi, \protect\BIBand{}
  Ye}]{agrawal2010correlation}
Agrawal S, Ding Y, Saberi A, Ye Y (2010) Correlation robust stochastic
  optimization. \emph{Proceedings of the twenty-first annual ACM-SIAM symposium
  on Discrete Algorithms}, 1087--1096 (SIAM).

\bibitem[{Agrawal et~al.(2020)Agrawal, Sethuraman, \protect\BIBand{}
  Zhang}]{agrawal2020optimal}
Agrawal S, Sethuraman J, Zhang X (2020) On optimal ordering in the optimal
  stopping problem. \emph{Proceedings of the 21st ACM Conference on Economics
  and Computation}, 187--188.

\bibitem[{Arnosti \protect\BIBand{} Ma(2021)}]{arnosti2021tight}
Arnosti N, Ma W (2021) Tight guarantees for static threshold policies in the
  prophet secretary problem. \emph{arXiv preprint arXiv:2108.12893} .

\bibitem[{Bansal et~al.(2012)Bansal, Gupta, Li, Mestre, Nagarajan,
  \protect\BIBand{} Rudra}]{bansal2012lp}
Bansal N, Gupta A, Li J, Mestre J, Nagarajan V, Rudra A (2012) When lp is the
  cure for your matching woes: Improved bounds for stochastic matchings.
  \emph{Algorithmica} 63(4):733--762.

\bibitem[{Beyhaghi et~al.(2021)Beyhaghi, Golrezaei, Leme, P{\'a}l,
  \protect\BIBand{} Sivan}]{beyhaghi2020improved}
Beyhaghi H, Golrezaei N, Leme RP, P{\'a}l M, Sivan B (2021) Improved revenue
  bounds for posted-price and second-price mechanisms. \emph{Operations
  Research} 69(6):1805--1822.

\bibitem[{Bradac et~al.(2019)Bradac, Singla, \protect\BIBand{}
  Zuzic}]{bradac2019near}
Bradac D, Singla S, Zuzic G (2019) (near) optimal adaptivity gaps for
  stochastic multi-value probing. \emph{Approximation, Randomization, and
  Combinatorial Optimization. Algorithms and Techniques (APPROX/RANDOM 2019)}
  (Schloss Dagstuhl-Leibniz-Zentrum fuer Informatik).

\bibitem[{Chen et~al.(2009)Chen, Immorlica, Karlin, Mahdian, \protect\BIBand{}
  Rudra}]{chen2009approximating}
Chen N, Immorlica N, Karlin AR, Mahdian M, Rudra A (2009) Approximating matches
  made in heaven. \emph{International Colloquium on Automata, Languages, and
  Programming}, 266--278 (Springer).

\bibitem[{Cominetti et~al.(2010)Cominetti, Correa, Rothvo{\ss},
  \protect\BIBand{} Mart{\'\i}n}]{cominetti2010optimal}
Cominetti R, Correa JR, Rothvo{\ss} T, Mart{\'\i}n JS (2010) Optimal selection
  of customers for a last-minute offer. \emph{Operations research}
  58(4-part-1):878--888.

\bibitem[{Esfandiari et~al.(2017)Esfandiari, Hajiaghayi, Liaghat,
  \protect\BIBand{} Monemizadeh}]{esfandiari2017prophet}
Esfandiari H, Hajiaghayi M, Liaghat V, Monemizadeh M (2017) Prophet secretary.
  \emph{SIAM Journal on Discrete Mathematics} 31(3):1685--1701.

\bibitem[{Fu et~al.(2018)Fu, Li, \protect\BIBand{} Xu}]{fu2018ptas}
Fu H, Li J, Xu P (2018) A ptas for a class of stochastic dynamic programs.
  \emph{45th International Colloquium on Automata, Languages, and Programming
  (ICALP 2018)} (Schloss Dagstuhl-Leibniz-Zentrum fuer Informatik).

\bibitem[{Gallego \protect\BIBand{} Segev(2022)}]{gallego22aconstructive}
Gallego G, Segev D (2022) A constructive prophet inequality approach to the
  adaptive probemax problem. \emph{arXiv preprint arXiv:2210.07556} .

\bibitem[{Gallego \protect\BIBand{} Topaloglu(2019)}]{gallego2019revenue}
Gallego G, Topaloglu H (2019) \emph{Revenue management and pricing analytics},
  volume 209 (Springer).

\bibitem[{Gandhi et~al.(2006)Gandhi, Khuller, Parthasarathy, \protect\BIBand{}
  Srinivasan}]{gandhi2006dependent}
Gandhi R, Khuller S, Parthasarathy S, Srinivasan A (2006) Dependent rounding
  and its applications to approximation algorithms. \emph{Journal of the ACM
  (JACM)} 53(3):324--360.

\bibitem[{Goodrich \protect\BIBand{} Tamassia(2001)}]{goodrich2001algorithm}
Goodrich MT, Tamassia R (2001) \emph{Algorithm design: foundations, analysis,
  and internet examples} (John Wiley \& Sons).

\bibitem[{Gupta \protect\BIBand{} Nagarajan(2013)}]{gupta2013stochastic}
Gupta A, Nagarajan V (2013) A stochastic probing problem with applications.
  \emph{International Conference on Integer Programming and Combinatorial
  Optimization}, 205--216 (Springer).

\bibitem[{Gupta et~al.(2016)Gupta, Nagarajan, \protect\BIBand{}
  Singla}]{gupta2016algorithms}
Gupta A, Nagarajan V, Singla S (2016) Algorithms and adaptivity gaps for
  stochastic probing. \emph{Proceedings of the twenty-seventh annual ACM-SIAM
  symposium on Discrete algorithms}, 1731--1747 (SIAM).

\bibitem[{Gupta et~al.(2017)Gupta, Nagarajan, \protect\BIBand{}
  Singla}]{gupta2017adaptivity}
Gupta A, Nagarajan V, Singla S (2017) Adaptivity gaps for stochastic probing:
  Submodular and xos functions. \emph{Proceedings of the Twenty-Eighth Annual
  ACM-SIAM Symposium on Discrete Algorithms}, 1688--1702 (SIAM).

\bibitem[{Hill(1983)}]{hill1983prophet}
Hill T (1983) Prophet inequalities and order selection in optimal stopping
  problems. \emph{Proceedings of the American Mathematical Society}
  88(1):131--137.

\bibitem[{Kleinberg \protect\BIBand{} Weinberg(2012)}]{kleinberg2012matroid}
Kleinberg R, Weinberg SM (2012) Matroid prophet inequalities. \emph{Proceedings
  of the forty-fourth annual ACM symposium on Theory of computing}, 123--136.

\bibitem[{Krengel \protect\BIBand{} Sucheston(1977)}]{krengel1977semiamarts}
Krengel U, Sucheston L (1977) Semiamarts and finite values. \emph{Bulletin of
  the American Mathematical Society} 83(4):745--747.

\bibitem[{Krengel \protect\BIBand{} Sucheston(1978)}]{krengel1978semiamarts}
Krengel U, Sucheston L (1978) On semiamarts, amarts, and processes with finite
  value. \emph{Probability on Banach spaces} 4:197--266.

\bibitem[{Mitzenmacher \protect\BIBand{}
  Upfal(2017)}]{mitzenmacher2017probability}
Mitzenmacher M, Upfal E (2017) \emph{Probability and computing: Randomization
  and probabilistic techniques in algorithms and data analysis} (Cambridge
  university press).

\bibitem[{Purohit et~al.(2019)Purohit, Gollapudi, \protect\BIBand{}
  Raghavan}]{purohit2019hiring}
Purohit M, Gollapudi S, Raghavan M (2019) Hiring under uncertainty.
  \emph{International Conference on Machine Learning}, 5181--5189 (PMLR).

\bibitem[{Samuel-Cahn(1984)}]{samuel1984comparison}
Samuel-Cahn E (1984) Comparison of threshold stop rules and maximum for
  independent nonnegative random variables. \emph{the Annals of Probability}
  1213--1216.

\bibitem[{Segev \protect\BIBand{} Singla(2021)}]{segev2021efficient}
Segev D, Singla S (2021) Efficient approximation schemes for stochastic probing
  and prophet problems. \emph{Proceedings of the 22nd ACM Conference on
  Economics and Computation}, 793--794.

\bibitem[{Yan(2011)}]{yan2011mechanism}
Yan Q (2011) Mechanism design via correlation gap. \emph{Proceedings of the
  twenty-second annual ACM-SIAM symposium on Discrete Algorithms}, 710--719
  (SIAM).

\end{thebibliography}

% FOR SUBMISSION
\ECSwitch
% \ECDisclaimer
% \ECHead{E-Companion}

% FOR SSRN
%\clearpage

% Appendix here
% Options are (1) APPENDIX (with or without general title) or
%             (2) APPENDICES (if it has more than one unrelated sections)
% Outcomment the appropriate case if necessary
%
% \begin{APPENDIX}{<Title of the Appendix>}
% \end{APPENDIX}
%
%   or
%

\begin{APPENDICES}
\crefalias{section}{appendix}
\crefalias{subsection}{appendix}

\DoubleSpacedXI
% \linespread{\appendixspace}\selectfont{}

\begin{Large}
\noindent Supplementary Material for ``Selection and Ordering Policies for Hiring Pipelines via Linear Programming''
\end{Large}

\crefalias{section}{appendix}
\crefalias{subsection}{appendix}

\section{Useful Known Results \label{sec:known_results}}

In this section, we provide some concepts and known results that will be used in our algorithms and analyses. We start by defining weighted rank functions for $k$-uniform matroids, which have a close connection to both the expected rewards of our algorithms and our LP benchmarks. We use correlation gaps for this family of functions to relate these two quantities. We also review the dependent rounding scheme developed by \citet{gandhi2006dependent}, which we use both in the implementation of our algorithms and as a tool for analysis. \orrev{We close by providing a closed-form formula for the expectation of a truncated Poisson random variable that is used throughout the paper.}

\subsection{Weighted Rank Functions and Correlations Gaps \label{app:weighted_rank}} Let \orrev{$(w_i)_{i\in [n]}$ be a vector of non-negative} weights and $D\subseteq [n]$. Define the weighted rank function for the $k$-uniform matroid $w^*:2^{[n]} \to [0,\infty)$ as \[w^*(D):=\max_{R\subseteq D, |R|\leq k} \sum_{i\in R} w_i.\]
Loosely speaking, the reward obtained by our algorithms can be expressed as weighted rank functions of $k$-uniform matroids. Here, $D$ is to be interpreted as the set of candidates eligible for hire, the weights are to be interpreted as the expected reward collected from each candidate conditional on being eligible, and the weighted rank function selects the $k$ highest rewards from the set $D$.

Let $D\sim\mathcal{D}$, with $\DD$ a distribution over $2^{[n]}$, and $q_i = \PP(i\in D)$. Let $\Tilde{D}\sim\mathcal{I}(\DD)$, where $\mathcal{I}(\DD)$ is a distribution over $2^{[n]}$ in which each $i\in D$ independently with probability $q_i$. The correlation gap of a set function $f:2^{[n]}\to [0,\infty)$ is defined by
\[ \sup_{\DD} \frac{\E_{D\sim \DD}(f(D))}{\E_{D\sim \mathcal{I}(\DD)}(f(D))}. \]
This concept was first formalized by \citet{agrawal2010correlation}. In words, it quantifies how much we can win by correlating the outcome of $\{i\in D\}$ for $i\in [n]$ while maintaining the marginal probabilities. We use a result from \citet{yan2011mechanism} concerning the correlation gap of weighted rank functions for $k$-uniform matroids.
\begin{proposition}[Lemma 4.4 from \citep{yan2011mechanism}]\label{prop:corrgap}
For any $n,k\geq 1$, the correlation gap of the weighted rank function of a $k$-uniform matroid of size $n$ is at most
 $(1- \frac{e^{-k}k^k}{k!})^{-1}$.
\end{proposition}

\subsection{GKPS Dependent Rounding \label{app:gkps_rounding}} \citet{gandhi2006dependent} developed a dependent rounding scheme that we use for establishing our results. This dependent rounding scheme receives as an input a bipartite graph $(A,B,E)$ with weights $x_{ij}\in [0,1]$ for all edges $(i,j)\in E$. The output is, for each edge, a \orrev{random variable} $X_{ij}\in\{0,1\}$. For any vertex $i\in A$, define the fractional degree $\ell_i = \sum_{j:(i,j)\in E} x_{ij}$. Analogously, for $j\in B$ define $\ell_j = \sum_{i:(i,j)\in E} x_{ij}$. The outputs satisfy the following three properties: 
\begin{itemize}
    \item[(P1)] Marginal distribution: $E(X_{ij})=x_{ij}$ for every $(i,j)\in E$,
    \item[(P2)] Degree preservation: with probability 1 it holds that $\sum_{j:(i,j)\in E} X_{ij} \in \{\lfloor \ell_i\rfloor,\lceil \ell_i\rceil \}$ for all $i \in A$ and $\sum_{i:(i,j)\in E} X_{ij} \in \{\lfloor \ell_j\rfloor,\lceil \ell_j\rceil \}$ for all $j \in B$, 
    \item[(P3)] Negative correlation: For any vertex $i\in A\cup B$, any subset $S$ of edges incident in $i$ and any $b\in\{0,1\}$, it holds that
    \[ \PP\left( \underset{e\in S}{\bigcap} \{X_e =b\}\right) \leq \underset{e\in S}{\prod} \PP(X_e =b). \]
\end{itemize}
For the special case when the bipartite graph is a star graph (i.e. $A=[n], |B|=1$) the input are simply weights $x_i\in[0,1]$ for all $i\in[n]$, and the negative correlation property can be stated as: for any subset $M\subseteq [n]$ and any $b\in\{0,1\}$ it holds that
    \begin{equation}
        \PP\left( \underset{i\in M}{\bigcap} \{X_i =b\}\right) \leq \underset{i\in M}{\prod} \PP(X_i =b). \nonumber
    \end{equation}

\subsection{\orrev{Expectation of Truncated Poisson Random Variable} \label{sec:TruncPois}}

We provide an elementary calculation used for deriving a closed-form formula for the expectation of a truncated Poisson random variable.
\begin{proposition}\label{prop:exppois}
For $k\in \mathbb{N}$, let $Z$ be a Poisson random variable with mean $k$. Then
    \[ \E(\min\{Z,k\}) = k\left(1 - e^{-k}\frac{k^k}{k!}\right). \]
\end{proposition}
\proof{Proof.} 
\begin{align*}
    \E(\min\{Z,k\}) &= \sum_{j=1}^k j \frac{e^{-k}k^j}{j!} + k\left(1 - \sum_{j=0}^k \frac{e^{-k}k^j}{j!} \right)\\
    &= k + k e^{-k} \left( \sum_{j=1}^k \frac{k^{j-1}}{(j-1)!} - \sum_{j=0}^k \frac{k^{j}}{j!} \right)\\
    &= k + k e^{-k} \left( \sum_{j=0}^{k-1} \frac{k^{j}}{j!} - \sum_{j=0}^k \frac{k^{j}}{j!} \right)\\
    &= k\left(1 - e^{-k}\frac{k^k}{k!}\right). \qed
\end{align*}

\section{Proofs of \Cref{sec:ptk} \label{ap:ptk}}

\subsection{Proof of \Cref{lem:upper_bound_PM} \label{sec:upper_bound_PM_pf}}
Consider any algorithm and let $Y_i$ be the indicator of applicant $i$ getting an interview and $X_{ij}$ the indicator of $i$ being hired and $V_i = r_j$. Also let $Q_{ij}$ be the indicator of $V_i = r_j$. We will show that $x_{ij}=\E(X_{ij})$ and $y_i = \E(Y_i)$ is feasible, and that the objective function is equal to the expected reward of the algorithm.

In order to hire an applicant $i$ with $V_i=r_j$, we need to interview the applicant and that the applicant has value $r_j$. This translates to $X_{ij} \leq Y_i Q_{ij} $, and by taking expectation and using that $Y_i$ and $Q_{ij}$ are independent we get constraint (\ref{const:xyq}). The algorithm can interview at most $T$ applicants, so $\sum_{i\in [n]} Y_i \leq T$. Again taking expectation we get constraint (\ref{const:yleqT}). The algorithm can hire at most $k$ applicants, so $\sum_{(i,j)\in [n]\times[J]} X_{ij}\leq k$, by taking expectation we get constraint (\ref{const: xleqk}). The remaining constraints are clearly satisfied.

Finally, we have that the reward of the algorithm equals $\sum_{(i,j)\in [n]\times[J]} X_{ij}r_j$, so the expected reward of the algorithm is equal to the objective function of the linear program.

\subsection{Proof of \Cref{lem:bfs2}\label{sec:bfs_pf}}
The linear program has $Jn+n$ variables and $2Jn + 2n +2$ constraints. This proof relies on two observations, the first one being that constraints $y_i\geq 0$ and $y_i \leq 1$ cannot be tight simultaneously. The second observation is that constraints $x_{ij} \geq 0$ and $x_{ij} \leq y_i q_{ij}$ cannot be tight simultaneously unless $y_i =0$. If that is the case, then we have that
 constraints
\[ y_i\geq 0,\quad x_{ij}\geq 0, \quad x_{ij} \leq q_{ij} y_i \]
are tight, but linearly dependent. With these two observations, we conclude that
\begin{itemize}
    \item[(i)] for each $i\in[n]$, we can only count one of $y_i\geq 0$ and $y_i \leq 1$ as a linearly independent tight constraint,
    \item[(ii)] for each $(i,j)\in[n]\times [J]$, we can only count one of $x_{ij} \geq 0$ and $x_{ij} \leq q_{ij} y_i$ as a linearly independent tight constraint.
\end{itemize}
As we need $Jn + n $ linearly independent tight constraints for $x,y$ to be a basic feasible solution and we only have two other constraints, we can drop at most 2 tight constraints out of the ones listed in points (i) and (ii). This shows that at most 2 components of $y$ are different from 0 or 1.

We still need to show that if there are two fractional components, they will add up to 1. This holds because if there are two fractional components, then constraint (\ref{const:yleqT}) is necessarily tight. This implies that $T-1$ components of $y$ are equal to 1, as setting $T-2$ or less would contradict the tightness of (\ref{const:yleqT}), and setting $T$ would contradict the two fractional components. Combining this with the tightness of (\ref{const:yleqT}) gives
\[ T =  \sum_{i\in [n]:y_i =1} y_i + \sum_{i\in [n]:0<y_i <1} y_i =  T-1 + \sum_{i\in [n]:0<y_i <1} y_i , \]
from which we conclude that
\[ \sum_{i\in [n]:0<y_i <1} y_i=1 .\]

\subsection{Proof of \Cref{lem:seq_exp_alg} \label{sec:seq_exp_alg_pf}}

When taking expectations we get
\begin{align*}
     \E\left(  V_i \Ind\{N_{i-1} < k\}  Z_i  \right)  & = \E\left(    V_i  | \Ind\{N_{i-1} < k\}Z_i =1 \right) \E(\Ind\{N_{i-1} < k\}Z_i)\\
     & = v_i\E(\Ind\{N_{i-1} < k\}Z_i).
\end{align*}
This expression holds because $\Ind\{N_{i-1} < k\}Z_i =1 $ is equivalent to applicant $i$ being hired and $v_i$ is the expected value of applicant $i$ given that they were interviewed and the algorithm would decide that they are hired:
\begin{align}
    \E(V_i|\Ind\{N_{i-1} < k\}Z_i=1) &=  \E(V_i|Z_i=1) \label{eq:line1} \\
    &= \sum_{j=1}^J r_{j} \PP(Q_{ij}=1|Z_i=1)\nonumber \\
    &= \sum_{j=1}^J r_{j} \PP(Z_i=1|Q_{ij}=1) \frac{\PP(Q_{ij}=1)}{\PP(Z_i=1)} \nonumber \\
    &= \sum_{j=1}^J r_{j} \PP(Z_i=1|Q_{ij}=1) \frac{q_{ij}}{p_iy_i}\label{eq:line4} \\
    &=\sum_{j=1}^J r_{j} \PP(Y_i=1|Q_{ij}=1)\PP(P_i=1|Q_{ij}=1) \frac{q_{ij}}{y_ip_i}\label{eq:line5} \\
    &=\sum_{j=1}^J r_{j} y_i \PP(P_i=1|Q_{ij}=1) \frac{q_{ij}}{y_ip_i} \label{eq:line6}\\
    &=\sum_{j=1}^J r_{j} y_i \frac{x_{ij}}{y_iq_{ij}} \frac{q_{ij}}{y_ip_i} \label{eq:line7}\\
    &= \frac{\sum_{j=1}^J r_i x_{ij}}{\sum_{j=1}^Jx_{ij} } = v_i. \label{eq:line8}
\end{align}
\orrev{In \orrev{equality \ref{eq:line1}} line we simply use that the value of an applicant is independent of the remaining positions left when they would be interviewed. In \orrev{equality \ref{eq:line4}}, we replaced the know probabilities in the fraction: $\PP(Q_{ij}=1)=q_{ij}$ and $\PP(Z_i=1) = \PP(P_i=1)\PP(Y_i=1) = p_iy_i$ since they are independent. In \orrev{equality \ref{eq:line5}}, we use that $Y_i$ and $P_i$ are independent. In \orrev{equality \ref{eq:line6}}, we replace $\PP(Y_i=1|Q_{ij}=1)= \PP(Y_i=1)=y_i$, since $Y_i$ is to be decided before knowing the realization of the values. In \orrev{equality \ref{eq:line7}} we replace $\PP(P_i=1|Q_{ij}=1)= x_{ij}/(y_iq_{ij})$. In \orrev{equality \ref{eq:line8}}, we simply replace the definition of $p_i$ and rearrange \orrev{the terms to} obtain the desired equality.}

With this equivalence, we get the following expression for the reward of $\ALG_\ptk$:
\[ \sum_{i=1}^n v_i \E(\Ind\{N_{i-1} < k\}Z_i). \]

\subsection{Proof of \Cref{lem:samplepath} \label{sec:samplepath_pf}}

The right-hand side of (\ref{eq:indepineq}) can be interpreted as the reward of a policy that samples from $\Tilde{Y}$ instead of $Y$ for deciding the applicants that will be interviewed. Notice that $\ALG_{\ptk}$ selects a set of $T$ applicants to interview, but the independent analog might select an order of $T+1$ applicants (if $\Tilde{Y}_{i_1}=\Tilde{Y}_{i_2}=1$). Define $V:=\sum_{i=1}^n v_i Z_i \Ind\{ N_{i-1} < k \}$ and $\Tilde{V}:=\sum_{i=1}^n v_i \Tilde{Z}_i \Ind\{ \Tilde{N}_{i-1} < k \}$. Let $\FF$ be the sigma algebra generated by $\{P_i\}_{i\in[n]}$ (i.e. all the randomness except $Y_{i_1}$, $\Tilde{Y}_{i_1}$, $Y_{i_2}$ and $\Tilde{Y}_{i_2}$, as the remaining $Y_i$ and $\Tilde{Y}_i$ are deterministic). 

For proving the lemma we show that $\E(V-\Tilde{V}|\FF)\geq0$, which implies the result by using the law of iterated expectations. We do this by an exhaustive sample path analysis establishing the inequality for all possible events in $\FF$. 

\orrev{\textbf{Case 1: $P_{i_1}=P_{i_2}=0$.}} The first thing to notice is that if $P_{i_1}=P_{i_2}=0$, then $V$ and $\Tilde{V}$ are identical. Indeed, \orrev{neither $i_1$ or $i_2$ will be hired}, and $Z_i=\Tilde{Z}_i$ for the remaining applicants. 

\orrev{\textbf{Case 2: $P_{i_1}=1, P_{i_2}=0$.}} If \orrev{this is the case}, then $V$ and $\Tilde{V}$ have the same expectation. Indeed, the values of $Y_{i_2}$ and  $\Tilde{Y}_{i_2}$ \orrev{will not} have any effect on the applicants hired by the algorithm. The only difference in the selection could be made by $Y_{i_1}$ and $\Tilde{Y_{i_1}}$, which have the same marginal distribution, so the expectation is equal. %The same argument works to establish that $V$ and $\Tilde{V}$ have the same expectation when $P_{j_1}=0$ and $P_{j_2}=1$.

\orrev{\textbf{Case 3: $P_{i_1}=0, P_{i_2}=1$.}} This case is symmetric to Case 2.

\orrev{\textbf{Case 4: $P_{i_1}=P_{i_2}=1$.}} To analyze this case, define $\phi := \{ i\in [n]\setminus\{i_1,i_2\}  : i <i_2, Z_{i}=1 \}$, so $|\phi|$ is the amount of applicants other than $i_1$ scheduled to get an interview before $i_2$ and would make the cut. Clearly $\phi\in \FF$. \orrev{We further distinguish 3 sub-cases, depending on the number of elements in $\phi$.}

\orrev{\textit{Subcase 4.1: $|\phi| \geq k$.}} In this case we have $\E((V-\Tilde{V})|\FF)=0$. Indeed, as $|\phi| \geq k$, then no matter what happens with applicant $i_1$, applicant $i_2$ will never get an interview. The only difference can be made by different values of $Y_{i_1}$ and $\Tilde{Y}_{i_1}$, whose marginal distributions are identical, so the expectation must also be.

\orrev{\textit{Subcase 4.2: $|\phi| =k-1$.}} \orrev{In this subcase}, $V$ will select everything in $\phi$ plus either $i_1$ or $i_2$. On the other hand, $\Tilde{V}$ might select only $i_1$, only $i_2$, both, or neither. In the last case, the last selected index in $\Tilde{V}$ will be $i':=\inf\{i > i_2: Z_i=1\}$. Notice that $i'$ might not exist, in that case, we say $i'=n+1$ and $v_{n+1}=0$. This way we write
\begin{align*}
     \E(V|P_{i_1}=P_{i_2}=1, \phi ,|\phi|=k-1)& = \sum_{i\in\phi} v_i + y_{i_1} v_{i_1} + y_{i_2} v_{i_2},\\
     \E(\Tilde{V}|P_{i_1}=P_{i_2}=1, \phi ,|\phi|=k-1) &= \sum_{i\in\phi} v_i + y_{i_1} v_{i_1} + (1-y_{i_1})(y_{i_2} v_{i_2} + (1-y_{i_2})v_{i'}),
\end{align*}
so the difference is
\begin{align*}
    \E(V - \Tilde{V} |P_{i_1}=P_{i_2}=1, \phi ,|\phi|=k-1) 
    &= y_{i_2} v_{i_2} - y_{i_2}( y_{i_2} v_{i_2} + y_{i_1}v_{i'} ) \\
    &\geq y_{i_2}(v_{i_2} - v_{i_2}) =0,
\end{align*}
as $y_{i_1}+ y_{i_2} = 1 $ and $v_{i_2}\geq v_{i'}$.

\orrev{\textit{Subcase 4.3: $|\phi|\leq k-2$}}. Define $\phi':=\{i\neq i_1, i_2: \sum_{j=1,j\neq i_1, i_2}^i Z_j \leq k -2\}$ to be the indices of applicants that will be certainly selected both in $V$ and in $\Tilde{V}$. Define $j_1:=\inf\{ i\in [n]\setminus\{i_1,i_2\}: Z_i =1, \sum_{j=1,j\neq i_1,i_2}^i Z_j = k-1 \}$. If the set defining $j_1$ is empty we again say $j_1=T+1$. Notice that $j_1 \geq i_2$ because $|\phi|\leq k-2$. Applicant $j_1$ will always be selected by $V$, as it will select the $k-2$ elements in $\phi'$, either $i_1$ or $i_2$, and $j_1$, which will be the next best applicant remaining. On the other hand, $\Tilde{V}$ will also select all applicants in $\phi'$, but the remaining two positions can be filled in more ways. It can be the case that $\Tilde{V}$ picks either $i_1$ or $i_2$ plus $j_1$, but it also might happen that it picks both $i_1$ and $i_2$ and not pick $j_1$, or it can also pick neither of $i_1$ and $i_2$ and pick instead $j_1$ plus another another which we define as $j_2:=\inf\{i> j_1: Z_i=1\}$. With this in mind, we can express
\begin{align*}
     \E(V|P_{i_1}=P_{i_2}=1, \phi' ,|\phi|\leq k-2)& = \sum_{i\in\phi'} v_i + y_{i_1} v_{i_1} + y_{i_2} v_{i_2} + v_{j_1},\\
     \E(\Tilde{V}|P_{i_1}=P_{i_2}=1, \phi' ,|\phi|\leq k-2) &= \sum_{i\in\phi'} v_i + y_{i_1}y_{i_2}(v_{i_1}+v_{i_2})  \\
     & + (y_{i_1})^2(v_{i_1} + v_{j_1})+ (y_{i_2})^2(v_{i_2} + v_{j_1})  + y_{i_1}y_{i_2}(v_{j_1}+v_{j_2})\\
     &= \sum_{i\in\phi'} v_i + y_{i_1} v_{i_1} + y_{i_2} v_{i_2} + y_{i_1} v_{j_1} + y_{i_2}v_{j_2}.
\end{align*}
The difference is
\begin{align*}
     \E(V - \Tilde{V}|P_{i_1}=P_{i_2}=1, \phi' ,|\phi|\leq k-2) = v_{j_1} - y_{i_1} v_{j_1} - y_{i_2}v_{j_2} \geq v_{j_1} - v_{j_1} =0,
\end{align*}

\subsection{Proof of \Cref{thm:probetopk} \label{sec:probetopk_pf}}
To show the result we first show how to express the expected reward of the independent version of our algorithm, $\ALG'_\ptk$, and the objective function of $\LP_\ptk$ as expectations of weighted rank functions.

For the first one, we use the interpretation of the algorithm collecting value $v_i$ when applicants make the cut. Let $M=\{i: \Tilde{Z_i}=1\}$. As the algorithm will interview applicants in decreasing order of $v_i$, we can express the expected reward as $R_{\ALG'_{\ptk}} =v^*(M)$. Notice that $\PP(i\in M) = \E(Z_i) = y_i p_i$.

For expressing the objective function of the linear program as a weighted rank function we rewrite
\[ \sum_{i=1}^n \sum_{j=1}^J r_{ij} x_{ij} = \sum_{i=1}^n y_i \frac{\sum_{j=1}^J r_{ij} x_{ij}}{\sum_{j=1}^J x_{ij}} \frac{\sum_{j=1}^J x_{ij}}{y_i} = \sum_{i=1}^n y_i v_i p_i. \]
This way, we can use the dependent rounding procedure presented in \citet{gandhi2006dependent} to generate indicators $W_i$ with probabilities $w_i = y_i p_i$. Constraint (\ref{const: xleqk}) can be re-written as
\[ \sum_{i=1}^n\sum_{j=1}^J x_{ij} = \sum_{i=1}^n y_i \sum_{j=1}^J \frac{x_{ij}}{y_i} = y_i p_i \leq k ,\]
so the dependent rounding procedure ensures that $D = \{ i\in[n]: W_i =1 \}$ has cardinality at most $k$. With this, we can use the linearity of the expectation to write
\[ \sum_{i=1}^n v_i y_i p_i = \sum_{i=1}^n v_i \PP(W_i=1)  = \max_{R\subseteq D, |R|\leq k} \sum_{i\in R} v_i = v^*(D).  \]
The second equality holds because $D$ has a cardinality of at most $k$.

With these expressions, for any instance $I\in\I_\ptk$ we can write
\[ \frac{R_{\ALG'_{\ptk}}(I)}{\LP_\ptk(I)} = \frac{v^*(M)}{v^*(D)} \geq 1- \frac{e^{-k}k^k}{k!}. \]
The inequality follows from \Cref{prop:corrgap} by \citet{yan2011mechanism}, since $M$ and $D$ have the same marginal distributions.

The last step is to invoke \Cref{lem:samplepath} for the inequality $R_{\ALG_\ptk}(I) \geq R_{\ALG'_\ptk}(I) $.

\subsection{Tightness of Guarantee \label{sec:tight}}

\orrev{\orrev{We provide an instance that shows} that the guarantee in \Cref{thm:probetopk} is tight. Indeed, construct the instance with $n$ applicants \orrev{with i.i.d~valuations. Each applicant will have $V_i = 1 $ with probability $k/n$ or $V_i=0$ with probability $1-k/n$. The implied support of the valuations' distributions is $\{r_0,r_1\} = \{0,1\}$.}
%identical candidates with weighted Bernoulli values. In particular, $r_i =1$ and $q_i = k/n$ for all $i$.
There is no time constraint for this algorithm, i.e. $T=n$. Denote the described instance by $I^\ptk_{UB}$.}

\begin{proposition}\label{thm:ptk_ub}
For any $\pi\in \Pi^\ptk$,
\[ R_\pi(I_{UB}^\ptk) \leq \left(1 - \frac{e^{-k}k^k}{k!}\right)\LP_{\ptk}(I_{UB}^\ptk). \]
\end{proposition}
\proof{Proof.}\orrev{ %Since we are working with an instance of weighted Bernoulli values we can work with $\LP_\seq$ without loss of generality. 
\orrev{An optimal solution for $\LP_\ptk(I_{UB}^\ptk)$ will have $y_i = 1$, $x_{i1}=k/n$ and $x_{i0}=0$ for all $i$, and attain an objective value of $k$. } On the other hand, \orrev{an optimal algorithm} will interview all applicants \orrev{one by one}, until $k$ of them have value 1 and are hired \orrev{or there are no applicants left to interview}. The expected reward of the algorithm will be $\E(\min\{k,B_{n,k}\})$, where $B_{n,k}\sim Bin(n,k/n)$. If we make $n$ go to infinity, the expected reward will converge to
\[ \E(\min\{\text{Poisson}(k),k\}) = k\left( 1- \frac{e^{-k}k^k}{k!} \right) \]
\orrev{by \Cref{prop:exppois} in \Cref{sec:TruncPois}.} \qed}

It is worth noting that this tightness is only with respect to $\LP_\ptk$. Indeed, $\ALG_\ptk$ is optimal when applied to $I^\ptk$.

\subsection{Proof of \Cref{lem:equivalent_sol_for_seq} \label{sec:equivalent_sol_for_seq_pf} }
\orrev{Let $x,y$ be an optimal solution of $\LP_\ptk$.}
%with $x_i < y_iq_i$ for some $i$.
Construct $\Tilde{x},\Tilde{y}$ with $\Tilde{x}_i = x_i$ and $\Tilde{y}_i = x_i/q_i$ for all $i\in [n]$. It is clear to see that $\Tilde{x},\Tilde{y}$ is feasible and that the value of the objective function remains unchanged.

% \subsection{Proof of Proposition \ref{thm:ptk_ub}}

\section{Proofs of \Cref{sec:paral} \label{ap:paral}}

\subsection{Proof of \Cref{lem:paral_ub} \label{sec:paral_ub_pf}}
For an arbitrary policy let $Y_{ij}$ be the indicator that candidate $i$ receives an offer for position $j$. Let $P_{ij}$ be the indicator that candidate $i$ would accept an offer for position $j$ should they receive one.

Any policy satisfies that for each position $j$, at most $T$ offers can be sent. In terms of the indicators:
\[\sum_{i=1}^n Y_{ij} \leq T,\]
so constraint (\ref{const:time_limit}) follows by taking expectation. Similarly, each candidate can receive at most one offer:
\[ \sum_{j=1}^k Y_{ij} \leq 1, \]
so constraint (\ref{const:offer_limit}) follows by expectation. Next, for each position $j$, at most one candidate can be hired. In terms of our indicators:
\[ \sum_{i=1}^n Y_{ij}P_{ij} \leq 1, \]
so constraint (\ref{const:acceptance_limit}) is obtained by taking expectation \orrev{and using that $P_{ij}$ and $Y_{ij}$ are independent. Constraints of the form $0\leq y_{ij} \leq 1$ are clearly satisfied for all $i\in [n]$ and $j\in[k]$.}
%the acceptance/rejection decisions are made before the offering rounds start.

Finally, the reward collected by the policy is
\[  \sum_{j=1}^k \sum_{i=1}^n v_{ij} Y_{ij} P_{ij}. \]
Again by taking expectation and using the independence of $Y_{ij}$ and $P_{ij}$ we obtain that the expected reward of the policy is equal to the objective function of the linear program.

\subsection{Proof of \Cref{lem:paral_indep} \label{sec:paral_indep_pf}}
For simplicity assume that $v_{1,j} \geq v_{2,j}\geq \cdots \geq v_{n,j}$. We can write
\[ v^*_j(D) = v_{1,j} + \sum_{i=2}^n (v_{i-1,j} - v_{
i,j})  \Ind\{ \cap_{\ell=1}^{i-1 }\{\ell \in D\} ) .\]
By taking expectation we get
\[ \E(v^*_j(D)) = v_{1,j} + \sum_{i=2}^n (v_{i,j} - v_{
(i+1),j})  \PP( \cap_{\ell=1}^{i-1 } \{\ell \in D\} ) .\]
Now, using the negative correlation property (P3) together with (P1) and $(v_{i-1,j} - v_{i,j})\leq 0 $ we can conclude
\begin{align*}
    \E(v^*_j(D)) &\geq v_{1,j} + \sum_{i=2}^n (v_{i,j} - v_{
(i+1),j})  \prod_{\ell=1}^{i-1 }  \PP(\{\ell \in D\} ) \\
& =  v_{1,j} + \sum_{i=2}^n (v_{i,j} - v_{
(i+1),j})  \prod_{\ell=1}^{i-1 }  \PP(\{\ell \in \Tilde{D}\} )\\
& = \E(v^*_j(\Tilde{D})).
\end{align*}

\subsection{Counterexample for \Cref{lem:paral_indep} if $k\geq2$ \label{app:counterexample}}

\begin{example} \label{eg:negCorrelWorseThanIndep}
Consider four elements $a$, $b$, $c$ and $d$, all with identical weights equal to 1. Consider the following rounding scheme: pick any subset from $\{ \{a\},\{b\},\{c\},\{d\},\{a,b,c\},\{a,b,d\},\{a,c,d\},\{b,c,d\}\}$ with equal probability. This implies that any element will be included in the set of eligible elements $D$ with probability 1/2. It is not hard to see that this rounding scheme satisfies (P3). \orrev{Indeed, for subsets of cardinality 1, the right-hand side of (P3) is 1/2. The left-hand side is also 1/2: it can be obtained by choosing the corresponding singleton, plus either of the three subsets of cardinality 3 that contain the element in question. By the same simple counting argument, the property will also be satisfied with equality for subsets of sizes 2 and 3. For the full set, however, the right-hand side of (P3) will be 1/16, while the left-hand side will be 0.}

The expected reward collected by choosing the $k=2$ highest weights with this rounding scheme is 3/2: with probability 1/2 we will have one element in the subset, and with probability 1/2 we will have three elements in the subset, of which we can only choose 2. However, if we formed the eligible subset by including each of the elements independently with probability 1/2, the reward collected would be 
\[ \E(\min\{B_{4,1/2},2\}) =1\cdot\frac{4}{16}+2\cdot(\frac{6}{16}+\frac{4}{16}+\frac{1}{16})=\frac{13}{8}> \frac{3}{2},\]
where $B_{4,1/2}$ refers to a Binomial random variable with 4 independent trials of probability 1/2.
\end{example}

\subsection{Proof of \Cref{lem:list_bound} \label{sec:list_bound_pf}}

Let $W_{ij}$ be the output of the \citet{gandhi2006dependent} dependent rounding with input $y_{ij}p_{ij}$. These indicators satisfy $\E(W_{ij}) = y_{ij}p_{ij}$ and $\sum_{i=1}^n W_{ij} \leq 1$ with probability 1, since $y$ satisfies constraint (\ref{const:acceptance_limit}). Let $M_j = \{ i : W_{ij} =1  \}$. We can then express
\[ L_j^* = \sum_{i=1}^n y_{ij}p_{ij} = \E(\max_{i\in M} v_{ij}) = \E(v_j^*(M)). \]
\orrev{The lemma follows from \Cref{prop:corrgap}, since $v_j^*$ is a weighted rank function for a 1-uniform matroid, and $\tilde{D}$ has the same marginal distributions as $M$, but the elements are sampled independently.}

\subsection{Tightness of Guarantee \label{sec:parTight}}

{In this section, we} show that the guarantee obtained in \Cref{thm:paral} is tight. Consider an instance with $k$ identical positions and $kT$ identical candidates. In particular, $v_{ij}=1$ and $p_{ij}=1/T$ for all $(i,j)\in [kT]\times[k]$. Denote this instance by $I_{UB}^\paral$.

\begin{proposition}\label{thm:paralUB}
For any $\pi\in\Pi^\paral$,
\[ R_\pi(I_{UB}^\paral) \leq \left(1 - \frac{1}{e}\right) LP_\paral(I_{UB}^\paral) \]
\end{proposition}
\proof{Proof.} A feasible solution for $\LP_\paral$ is to set $y_{ij} = 1/k$ for all $(i,j) \in [kT]\times[k]$, achieving an objective of $k$. \orrev{On the other hand, \orrev{since all candidates are identical, forming} $k$ parallel lists with $T$ candidates each, and running each list in parallel \orrev{is clearly optimal}.} The reward collected by a single list is
\[ \E(\min\{ B_{T,1/T}, 1 \}) \to_{T\to\infty} \E(\min\{\text{Poisson}(1),1\}) = 1- \frac{1}{e}, \]
where $B_{T,1/T} \sim Bin(T,1/T)$. \orrev{The last equality is obtained by applying \Cref{prop:exppois} in \Cref{sec:TruncPois}.} Since there are $k$ identical lists, the ratio between the performance of the algorithm and the linear program is exactly $(1-1/e)$. \qed

\subsection{Proof of \Cref{thm:cost_of_batching} \label{sec:cost_of_batching_pf}}
Let $y$ be a solution for $\LP_\seq$. Let us construct $y'$, a solution for $\LP_\paral$. In particular, we let $y'_{ij} = y_i/k$ for all $(i,j) \in [n]\times [k]$. It is straightforward to see that the solution is feasible and that the values of the objective functions are equal, so $\LP_\paral(I) \geq \LP_\seq(I')$. For the other direction, let $z$ be a solution for $\LP_\paral$. We can construct $z'$ by setting $z'_{i} = \sum_{j=1}^k z_{ij}$. It is again straightforward that this solution is feasible in $\LP_\seq$ and that the values of the objective functions are equal, so so $\LP_\paral(I) \leq \LP_\seq(I')$.

% We again remark that $\ALG_\paral$ is actually optimal for $I_{UB}^\paral$.

% \subsection{Proof of Proposition \ref{thm:paralUB}}

\section{Proofs and Complementary Material of \Cref{sec:simul} \label{ap:simul}}

\subsection{\orrev{Computation of the Optimal Value-ordered Policy \label{app:simul_DP}} }

In this section, we explain how to obtain the optimal value-ordered policy for the Simultaneous Offering problem. The first step is to compute the distribution of the number of candidates that accept an offer, when the set of candidates that receive an offer is $S=\{1,\dots,m\}$, for all possible values of $m\in[n]$. For this, we use a dynamic program. For $j\leq m$, let $Q(j,m)$ be the probability that $j$ candidates accept an offer when candidates $1$ through $m$ get offers. We can obtain these values with $O(n^2)$ computations by solving the following system of equations using recursion:
\begin{align*}
    Q(1,1) &= p_1 &\\
    Q(0,1) &= 1-p_1&\\
    Q(j,m+1) &= p_{m+1}V(j-1,m) + (1-p_{m+1})V(j,m)&\quad \forall m\in[n-1],\quad j \leq m+1.
\end{align*}
Now let $V(m)$ denote the expected reward for sending offers to candidates in $S=\{1,\dots,m\}$. We can now compute
\[ U(m) = \sum_{i=1}^m v_i p_i - \sum_{j=k+1}^m (j-k)\cdot Q(j,m). \]
The last step is to pick $m^* \in \arg\max\{ U(m): m\in[n] \}$. Thus, we can find the optimal value-ordered policy in runtime $O(n^2)$.

\subsection{Example Illustrating Expected-value-ordered Policies' Bad Performance \label{app:EO_example} }

\begin{example} \label{eg:sim_EV_ordered}
Consider an instance with $k=1$ positions to fill {and} $n+1$ candidates. Candidate 1 has $v_1 = 1/n$ and $p_1 = 1$. Candidates $i=2,\dots, n+1$ have $v_i = 1 - 1/n$ and $p_i = 1/n$. Candidate 1 has a higher expected value than any other candidate, so any expected-value-ordered policy must include her. Since $p_1=1$, including any other candidate can only decrease the collected reward, so the optimal expected-value-ordered policy is to only send an offer to candidate 1 and obtain a reward of $1/n$ which vanishes as $n$ grows large. On the other hand, a valid policy is to send an offer to all candidates in $\{2,\dots,n+1\}$. Let $Z\sim \bin(n,1/n)$. The expected reward of this policy is
\begin{align*}
    n\left(1 - \frac{1}{n}\right) \frac{1}{n} - \E\left[\max\{Z-1,0\}\right]& = \left(1 - \frac{1}{n}\right) - \sum_{i=2}^n (i-1)\PP(Z=i)\\
    & = \left(1 - \frac{1}{n}\right) - \left[\sum_{i=2}^n i\PP(Z=i) - \sum_{i=2}^n \PP(Z=i) \right]\\
    & = \left(1 - \frac{1}{n}\right) - \left[ (\E(Z) - \PP(Z=1) ) - (1 - \PP(Z=1) - \PP(Z=0))  \right]\\
    &= 1 - \frac{1}{n} - \left( 1 - \frac{1}{n}\right)^n \underset{n\to\infty}{\to} 1 - \frac{1}{e}.
\end{align*}
This shows that for $n$ large enough, the ratio between the optimal expected-value-ordered policy and another policy can be made arbitrarily close to 0.
\orrev{Note that the greedy policy would also add candidate 1 first, since candidate 1 has the highest marginal benefit when the offer set is initially empty, hence the greedy policy would also perform as poorly as expected-value-ordered policies.}
\end{example}

\subsection{Proof of \Cref{lem:sim_LP_ub} \label{sec:sim_LP_ub_pf}}
By noting that {$f(x) = \max\{x,a\}$ for some constant $a$} is a convex function, we can apply Jensen's inequality to bound the objective function in $\OPT_\simul$. For any policy, let $Y_i$ denote the indicator that candidate $i$ {receives} an offer, and let $y_i = \E(Y_i)$. Similarly, let $P_i$ be the indicator that candidate $i$ would accept an offer if they receive one, with $\E(P_i)=p_i$. We can write the expected reward obtained by the policy as
\begin{align}
    &\E\left( \sum_{i \in [n]} v_i Y_i P_i  - \max\left\{ \sum_{i \in [n]}Y_i P_i -k, 0 \right\}\right)\nonumber \\
    \leq & \sum_{i\in [n]} v_i\E(Y_i P_i) - \max\left\{ \sum_{i\in[n]} \E(Y_i P_i) -k ,0 \right\} \label{eq:lem10-1} \\
    =& \sum_{i\in [n]} v_ip_iy_i - \max\left\{ \sum_{i\in[n]} p_i y_i -k ,0 \right\}\label{eq:lem10-2}  \\
    =& \sum_{i\in [n]} v_ip_iy_i + \min\left\{k - \sum_{i\in[n]} p_i y_i  ,0 \right\}\label{eq:simUB}
\end{align}
where in inequality \ref{eq:lem10-1} we use Jensen's inequality and in equation \ref{eq:lem10-2} we use the fact that $Y_i$ and $P_i$ are independent. We know that for any instance $I\in \I_\simul$, and any policy $\pi \in \Pi^\simul$, $R_\pi(I)$ is upper-bounded by \orrev{the expression in equation} \ref{eq:simUB}, so by optimizing this expression \orrev{over randomized policies, represented by $\{y_i\}_{i\in[n]}$,} we obtain an upper bound on $\OPT_\simul(I)$. \orrev{We can represent this optimization problem as a linear program by introducing auxiliary variable $z\in \R$, which will take the value of $\min\left\{k - \sum_{i\in[n]} p_i y_i,0 \right\} $. This can be easily accomplished by requiring $z\leq 0$ and $z \leq k - \sum_{i\in[n]} p_i y_i $. Since we are maximizing and $z$ appears with a positive sign in equation \ref{eq:simUB}, one of the two mentioned constraints will always be binding.}

\subsection{Proof of \Cref{lem:simul_mass_solution} \label{sec:simul_mass_solution_pf}}

We prove both statements by guessing a solution and showing that any other solution with the structure given in \Cref{lem:simDecreasingSol} achieves either the same or a worse objective value. Notice that this solution structure implies that the whole solution is completely determined by its total mass. In all cases we can assume that $z=\min\{0, k - \sum_{i\in [n]} y_i p_i\}$.

We start with the case $P_\high>k$. Let $j$ be the last index of $V_\high$. This is the same as saying that $j+1$ is the first index satisfying $v_{j+1}<1$. We can modify the guessed solution in two ways: by either increasing or decreasing the total mass. This first translates to either decreasing $y_j$ or increasing $y_{j+1}$. Since our guessed solution satisfies $\sum_{i\in[n]}y_i p _i = P_\high>k$, then $z>0$. If we decrease $y_j$ by $\delta>0$, the change in the objective function is upper-bounded by $ \delta p_j(1 - v_{j}) \leq 0 $, so the objective does not improve. The same will happen with each index $i<j$, the change in the objective function will be upper-bounded by $\delta p_i( 1- v_i)\leq 0$, so reducing the mass of our guessed solution cannot improve the objective value. On the other hand, if we increase $y_{j+1}$ by $\delta>0$, because  $z>0$ and it will not decrease, the change in the objective function will be $\delta p_{j+1}(v_{j+1}-1)\leq 0$. If we further increase indices $i>j+1$ by $\delta>0$, the change in the objective value will be $\delta p_i(v_i-1)\leq 0$. This way, cannot improve the objective value of our guessed solution by increasing its total mass and is therefore optimal.

The case when $P_\high<k$ follows the same logic. First, if $P_\tot<k$, then we can make $y_i =1$ for all $i\in [n]$ without incurring in any cost (since the total mass will always be less than $k$ and therefore $z=0$). If $P_\tot>k$, then our guessed solution will have mass $k$ and $z=0$. Let $j$ be the last index in our guessed solution such that $y_j>0$. If we reduce the mass of the solution by $0<\delta\leq y_j$, we will lose $p_j y_j$ and we will not reduce any cost since $z$ is already 0. The same will happen if we further decrease the mass from indices $i<j$. If we increase the mass of the solution by $\delta<1-y_j$, then $z$ will increase by the same amount and the change in the objective value will be $\delta p_j (v_j-1) \leq 0$. The same happens if we further increase the mass. We conclude that our guessed solution is also optimal in this case.

\subsection{Proof of \Cref{lem:simul_fixed_s} \label{sec:simul_fixed_s_pf}}

Let $y$ be an optimal solution of $\LP_\simul(I)$ and $y'$ be the alternative solution constructed by $\ALG_\simul^s$. Let $M = \sum_{i\in [n]} y_ip_i$, which is at most $k$ by \Cref{lem:simul_mass_solution}. By construction, $\sum_{i\in[n]}y_i'p_i = sM$. Let $\{D_i\}_{i=1}^n$ and $\{D_i'\}_{i=1}^n$ be collections of independent Bernoulli random variables with parameters $y_ip_i$ and $y_i' p_i$, respectively. Since $\ALG_\simul^s$ independently sends offers to candidates with probabilities $y_i'$, we interpret $D_i'$ as the indicator that candidate $i$ is hired. We can write
\begin{align*}
    R_{\ALG_\simul^s}(I) &= \sum_{i\in[n]} v_i \E(D_i') - \E\left(\max\left\{\sum_{i\in[n]}D_i'-k,0\right\}\right)\\
    &=  \sum_{i\in[n]} v_i \E(D_i') -  \sum_{i\in[n]} \E(D_i') + \E\left(\min\left\{\sum_{i\in[n]}D_i',k\right\}\right)\\
    & \geq \sum_{i\in[n]} v_i \E(D_i') -  \sum_{i\in[n]} \E(D_i') + \frac{\E\left(\min\left\{ \poiss(sM), k \right\} \right)}{sM}\sum_{i\in[n]} \E(D_i')\\
    & = \sum_{i\in[n]} v_i \E(D_i') +  \sum_{i\in[n]} \E(D_i') \left( -1 + \frac{\E\left(\min\left\{ \poiss(sM), k \right\} \right)}{sM} \right).
\end{align*}
In the inequality, we use a folklore splitting argument \orrev{(see, e.g., Lemma 4.2 in \citet{yan2011mechanism})} saying that if $\sum_{i\in[n] }\E(D_i')\leq sM$, then 
\[ \frac{\E\left(\min\{\sum_{i\in[n] }\E(D_i'),k\}\right)}{\sum_{i\in[n] }\E(D_i')} \geq  \frac{\E(\min\{\poiss(sk),k\})}{sM}. \]
The splitting argument uses the fact that $\bE[\min\{\poiss(x),k\}]/x$ is decreasing in $x$. With this last fact combined with  \Cref{lem:simul_mass_solution}, which gives us that $M\leq k$, we can further bound
%\wnote{I checked this, perfect!  Should also explicitly say that this argument depends on the fact that the function $\bE[\min\{\poiss(x),k\}]/x$ is decreasing in $x$ (you use this fact again below)}
%\wnote{Please check throughout the manuscript that you are using consistent notation for Poisson random variables}
\begin{equation}
     R_{\ALG_\simul^s}(I) \geq \sum_{i\in[n]} v_i \E(D_i') +  \sum_{i\in[n]} \E(D_i') \left( -1 + \frac{\E\left(\min\left\{ \poiss(sk), k \right\} \right)}{sk} \right). \label{eq:simul_alg_bound}
\end{equation}

Since $\sum_{i\in V_\high} p_i \leq k$, \Cref{lem:simul_mass_solution} gives us that $z=0$, thus we can write
\[ \LP_\simul(I) = \sum_{i\in [n]} v_i y_i p_i = \sum_{i\in[n]} v_i \E(D_i).  \]
By combining this with {equation} \ref{eq:simul_alg_bound} we can further bound
\begin{align}
     \frac{R_{\ALG_\simul^s}(I)}{\LP_\simul(I)} &\geq \frac{\sum_{i\in[n]} v_i \E(D_i')}{\sum_{i\in[n]} v_i \E(D_i)} +  \frac{\sum_{i\in[n]} \E(D_i')}{\sum_{i\in[n]} v_i \E(D_i)} \left( -1 + \frac{\E\left(\min\left\{ \poiss(sk), k \right\} \right)}{sk} \right) \nonumber\\
     & \geq s + \frac{\sum_{i\in[n]} D_i'}{ \sum_{i\in[n]}v_i D_i} \left( -1 + \frac{\E\left(\min\left\{ \poiss(sk), k \right\} \right)}{sk} \right) \label{eq:lem13-1}\\
     & \geq s + \frac{\sum_{i\in[n]} D_i'}{\tau \sum_{i\in[n]} D_i} \left( -1 + \frac{\E\left(\min\left\{ \poiss(sk), k \right\} \right)}{sk} \right)\label{eq:lem13-2}\\
     & = s + \frac{s}{\tau}\left(-1 + \frac{\E\left(\min\left\{ \poiss(sk), k \right\} \right)}{sk}\right).\label{eq:lem13-3}
\end{align}
\orrev{In inequality \ref{eq:lem13-1} we use that $v_i$ are decreasing in $i$, and since the solution $y'$ is constructed by truncating $y$, so
\[\frac{\sum_{i\in[n]} v_i \E(D_i')}{\sum_{i\in[n]} v_i \E(D_i)} \geq \frac{\sum_{i\in[n]}  \E(D_i')}{\sum_{i\in[n]}  \E(D_i)} = s.\]
In inequality \ref{eq:lem13-2}, we use that
 \[ -1 + \frac{\E\left(\min\left\{ \poiss(sk), k \right\} \right)}{sk} \leq -1 + \frac{\E\left(\poiss(sk) \right)}{sk} = 0, \]
 so the whole second term is non-positive and can only be reduced by bounding $v_i$'s by $\tau$ in the denominator. In \orrev{equality \ref{eq:lem13-3}}, we simply use the construction of $y'$ which scales the total mass of $y$ by a factor $s$.}

\subsection{Proof of \Cref{lem:simul_easy_case} \label{sec:simul_easy_case_pf}}

Let $I=(k,n,p,v)\in \I_\simul^\tau$ be such that $\sum_{i\in V_\high} p_i  = m> k$. Construct $I^{\ALT} = (k,n,p^\ALT,v)$ by setting $p_i^\ALT = p_i k/m$. This way $\sum_{i\in V_\high} p_i^\ALT = k $. We first show that for both of these instances, $s^*=1$. Since the optimal solutions of both linear programs will only have positive components for candidates in $V_\high$, we can without loss of generality assume that the instances only contain the candidates in $V_\high$. This can be easily seen by inspecting the expression
\[s - \frac{s}{\tau} + \frac{\E[\min\{\poiss(sk),k\}]}{\tau k}.\]
Since $\tau > 1$, the expression is clearly increasing in $s$, thus setting $s^*=1$ is optimal.

Now let $D_i$ be Bernoulli variables with parameters $p_i$ and $D_i^\ALT$ be Bernoulli random variables with parameters $p_i^\ALT$. For the original instance, we can write
\[ R_{\ALG_\simul^1}(I) = \sum_{i \in V_{\high}} v_i p_i - \E\left( \max\{ \sum_{i\in V_\high} D_i -k,0 \} \right) \]
and
\[ \LP_\simul(I) =\sum_{i \in V_{\high}} v_i p_i - (m-k). \]
For the alternative instance, we can also write
\[ R_{\ALG_\simul^1}(I^\ALT) = \sum_{i \in V_{\high}} v_i p_i \frac{k}{m} - \E\left( \max\{ \sum_{i\in V_\high} D_i^\ALT -k,0 \} \right) \]
and 
\[ \LP_\simul(I^\ALT) =\sum_{i \in V_{\high}} v_i p_i \frac{k}{m}. \]
We will show that
\begin{equation}
     R_{\ALG_\simul^1}(I) - R_{\ALG_\simul^1}(I^\ALT) \geq \LP_\simul(I) - \LP_\simul(I^\ALT), \label{eq:easy_case_lemma}
\end{equation}
which will imply the result.

\Cref{eq:easy_case_lemma} is equivalent to \[ m-k \geq  \E\left( \max\{ \sum_{i\in V_\high} D_i -k,0 \} \right) - \E\left( \max\{ \sum_{i\in V_\high} D_i^\ALT -k,0 \} \right). \]
The right-hand side of this inequality can be re-written as
\begin{align*}
    &\left[\sum_{i\in V_\high} \E(D_i) +  \E\left( \max\{ -k,- \sum_{i\in V_\high} D_i \} \right)   \right] - \left[\sum_{i\in V_\high} \E(D^\ALT_i) +  \E\left( \max\{ -k,- \sum_{i\in V_\high} D^\ALT_i \} \right)   \right]\\
    =& m - k + \left[ \E\left( \max\{ -k,- \sum_{i\in V_\high} D_i \} \right) - \E\left( \max\{ -k,- \sum_{i\in V_\high} D^\ALT_i \} \right) \right].
\end{align*}
Finally, $\E\left( \max\{ -k,- \sum_{i\in V_\high} D^\ALT_i \} \right)$ stochastically dominates $\E\left( \max\{ -k,- \sum_{i\in V_\high} D_i \} \right) $, so
\[ \E\left( \max\{ -k,- \sum_{i\in V_\high} D_i \} \right) - \E\left( \max\{ -k,- \sum_{i\in V_\high} D^\ALT_i \} \right) \leq 0, \]
which concludes the proof.

\subsection{Proof of \Cref{thm:simul_upper_bound} \label{sec:simul_upper_bound_pf}}

Recall that the instance has $n = j^2 + k$ candidates, with $j$ being a large integer. The first $j^2$ candidates are of type 1 and have $p_i = k/j$ and $v_i= \tau + \delta$, where $\delta = \varepsilon  \tau / j$ is small. The remaining $k$ candidates are of type 2, who have $p_i=1$ and $v_i = \tau$.

We first note that a feasible policy is to send offers to all type 2 candidates, which yields a reward of $k\tau$. Thus, $\OPT_\simul(I) \geq k\tau$. 

Now we show that an optimal value-ordered policy will never send offers to candidates of type 2 and that the number of offers sent to candidates of type 1 cannot be very large with respect to $j$. To make this formal, consider a value-ordered policy that sends offers only to candidates of type 1. The policy must decide the amount of type 1 candidates who are going to receive offers. Let $m_j$ be the optimal number of offers to be sent to these type of candidates. 
% \wnote{To make the definition of the sequence $(s_j)_j$ more formal, maybe we want to define $m_j$ to be an \textit{optimal} number for each $j$, and $s$ to be the $\liminf$?  And then we suppose for contradiction that $s=\infty$.}
Notice that $0\leq m_j \leq j^2$.  Define $s_j:=m_j/j$, $s = \lim_{j\to\infty} s_j$. We will show that $s<\infty$, since having $s_j\to\infty$ as $j\to\infty$ would imply that the policy has a reward that diverges to $-\infty$. Let $A_j$ denote the event $\{\bin(m_j,k/j)>k\}$. The reward of a policy that sends offers to $m_j$ type 1 candidates when there are $j^2$ of them available is
\begin{align}
    \ALG_{s_j} &=(\tau + \delta)\E(\bin(m_j,k/j)) + \E(\min\{0, k - \bin(m_j,k/j)  \})\nonumber\\
    &= (\tau + \delta)k\frac{m_j}{j} + \E(\min\{0, k - \bin(m_j,k/j) \})\label{eq:simul_resume}\\
    & = (\tau + \delta)k\frac{m_j}{j} + \E(k-\bin(m_j,k/j)|A_j)\PP(A_j)\nonumber\\
    & \leq k + (\tau + \delta)k\frac{m_j}{j}  - \E(\bin(m_j,k/j))\PP(A_j)\nonumber\\
    &= k + k\frac{m_j}{j}(\tau + \delta - \PP(A_j)).\nonumber
\end{align}
In the inequality we use the fact that we are subtracting the Binomial random variable in the expectation, so removing the conditional $\{\bin(m_j,k/j)>k\}$ can only make our expression larger. Now suppose that $m_j/j \to \infty$ as $j$
 grows to infinity. That implies that the mean of $\bin(m_j,k/j)$ also diverges, so $\PP(A_j) = \PP(\bin(m_j,k/j)>k)$ grows to 1. This in turn means that for $j$ large enough, $(\tau + \delta - \PP(A_j))<0$, so the expected reward is upper-bounded by a sequence that diverges to $-\infty$. This has two implications: first, an optimal value-ordered policy will never send offers to candidates of type 2, since they can only add at most a constant reward for being included. Second, for upper-bounding the expected reward of any value-ordered policy in this instance, we can assume without loss of generality that $s< \infty$.
 
 To establish our upper bound let us resume from \Cref{eq:simul_resume} and develop
 \begin{align}
     \ALG_{s_j} &= (\tau + \delta)k\frac{m_j}{j} + \E(\min\{0, k - \bin(m_j,k/j) \})\nonumber\\
     &=(\tau+\delta)k\frac{m_j}{j} - \E(\bin(m_j,k/j)) + \E(\min\{k, \bin(m_j,k/j)\}) \label{eq:thm4-1}\\
    &=\tau\frac{m_jk}{j} + \frac{\varepsilon k\tau m_j}{j^2} - \frac{m_jk}{j} + \E(\min\{k, \bin(m_j,k/j)\})\label{eq:thm4-2}\\
     &  \leq \varepsilon k \tau +k\frac{m_j}{j} \left(  \tau - 1 \right) + \E(\min\{k, \bin(m_j,k/j)\}) \label{eq:thm4-3}\\
    & \underset{j\to\infty}{\to} \varepsilon k \tau + \tau k s  - sk + \E(\min\{\poiss(sk),k\}). \label{eq:thm4-4}
 \end{align}
\orrev{In the \orrev{equation \ref{eq:thm4-1}}, we rearrange terms. In the \orrev{equation \ref{eq:thm4-2}}, we replace the definition of $\delta$, the expectation of the Binomial random variable, and rearrange terms. In \orrev{equation \ref{eq:thm4-3}}, we use the fact that $m\leq  j^2$ and $\varepsilon>0$. In the \orrev{equation \ref{eq:thm4-4}} we use the fact that $\lim_{j\to\infty}s_j <\infty$, so $\bin(m,k/j)$ converges in distribution to $\poiss(sk)$ and $\min\{x,k\}$ is a continuous bounded function.}

% \begin{align}
%     \ALG_{s_j} &=(\tau + \delta)\E(\bin(m_j,k/j)) + \E(\min\{0, k - \bin(m_j,k/j)  \})\nonumber\\
%     &=(\tau+\delta)\frac{mk}{j} - \E(\bin(m,k/j)) + \E(\min\{k, \bin(m,k/j)\}) \nonumber\\
%     &=\tau\frac{m_jk}{j} + \frac{\varepsilon k\tau m_j}{j^2} - \frac{m_jk}{j} + \E(\min\{k, \bin(m_j,k/j)\})\nonumber\\
%      &  \leq \varepsilon k \tau +k\frac{m_j}{j} \left(  \tau - 1 \right) + \E(\min\{k, \bin(m_j,k/j)\}) \label{eq:simul_alg_ub}\\
%     & \underset{j\to\infty}{\to}\tau k s + \varepsilon k \tau - sk + \E(\min\{\poiss(sk),k\}).
% \end{align}

Putting all together,
we obtain that
\[ \frac{\ALG_s}{\OPT_\simul(I)} \leq \varepsilon + s - \frac{s}{\tau} + \frac{ \E(\min\{\poiss(sk),k\})}{k\tau}. \]
The proof is concluded by taking the supremum over $s$.

% \subsection{Proof of \Cref{prop:simul_foc}}

% \subsection{Proof of \Cref{prop:simul_tight}}

\subsection{\orrev{Tightness Analysis} \label{app:simil:plots} }

\orrev{In this section, we provide conditions for our derived bounds to be tight.} Notice that the only difference between the optimization problems defining $\alpha^\tau_k$ and $\beta^\tau_k$ is that the former has the constraint $s\leq 1$.
%restricts the parameter $s$ to be at most 1.
Neither set of constants has a closed-form value, but since their objective functions are concave, we can characterize their optimal solutions using first-order conditions.

\begin{proposition}\label{prop:simul_foc}
For $\tau \in (0,1)$, let $s^*_\alpha$ and $s^*_\beta$ be optimal solutions for $\alpha_k^\tau$ and $\beta_k^\tau$, respectively. Then $s^*_\alpha = \min\{s^*_\beta,1\}$ and $s^*_\beta$ is the solution of
\[ \tau = 1 - e^{-sk}\sum_{j=0}^{k-1}\frac{(s k)^j}{j!}. \]

\end{proposition}
\proof{Proof.} We first prove the characterization of $s^*_\beta$. For that, we show that the objective function of the optimization problems is concave. Let us develop:
\begin{align*}
s - \frac{s}{\tau} + \frac{\E[\min\{\poiss(sk),k\}]}{k \tau }=&s - \frac{s}{\tau} + \frac{1}{k\tau}\left[ \sum_{j=0}^{k-1}j\frac{e^{-sk}(sk)^j}{j!} + k \left(1 - \sum_{j=0}^{k-1}\frac{e^{-sk}(sk)^j}{j!}\right) \right]\\
    &= s - \frac{s}{\tau} + \frac{1}{\tau} - \frac{1}{k\tau} \sum_{j=0}^{k-1} (k-j)\frac{e^{-sk}(sk)^j}{j!}.
\end{align*}
By taking derivative with respect to $s$ we get
\begin{align}
    & 1 - \frac{1}{\tau} - \frac{1}{k\tau}\sum_{j=0}^{k-1}\frac{(k-j)}{j!}\left[ \frac{je^{-sk}(sk)^j}{s} - k e^{-sk}(sk)^j \right] \nonumber\\
    =& 1 - \frac{1}{\tau} - \frac{e^{-sk}}{\tau}\sum_{j=1}^{k-1}(k-j)\frac{(sk)^{j-1}}{(j-1)!} + \frac{e^{-sk}}{\tau}\sum_{j=0}^{k-1}(k-j)\frac{(sk)^{j}}{(j)!}\nonumber \\
    =& 1 - \frac{1}{\tau} - \frac{e^{-sk}}{\tau}\sum_{j=0}^{k-2}(k-(j+1))\frac{(sk)^{j}}{j!} + \frac{e^{-sk}}{\tau}\sum_{j=0}^{k-1}\frac{(k-j)}{(j)!}(sk)^{j} \nonumber\\
     =& 1 - \frac{1}{\tau} - \frac{e^{-sk}}{\tau}\sum_{j=0}^{k-2}(k-(j+1))\frac{(sk)^{j}}{j!} + \frac{e^{-sk}}{\tau}\sum_{j=0}^{k-1}\frac{(k-j)}{(j)!}(sk)^{j}\nonumber\\
     =& 1 - \frac{1}{\tau} + \frac{e^{-sk}}{\tau}\sum_{j=0}^{k-2}\frac{(sk)^{j}}{j!} - \left[ \frac{e^{-sk}}{\tau}\sum_{j=0}^{k-2}(k-j)\frac{(sk)^{j}}{j!} - \frac{e^{-sk}}{\tau}\sum_{j=0}^{k-1}\frac{(k-j)}{(j)!}(sk)^{j}\right]\nonumber\\
     =& 1 - \frac{1}{\tau} + \frac{1}{\tau}\sum_{j=0}^{k-1}e^{-sk}\frac{(sk)^{j}}{j!}. \label{eq:1st_derivative}
\end{align}
The second derivative of the objective function is
\begin{align*}
    & \frac{1}{\tau} \left[ \sum_{j=0}^{k-1}\frac{je^{-sk}(sk)^j}{j!s} -  \sum_{j=0}^{k-1} k e^{-sk}(sk)^j \right]\\
    = & \frac{1}{\tau} \left[ \sum_{j=1}^{k-1}\frac{e^{-sk}(sk)^j}{j!s} -  \sum_{j=0}^{k-1}\frac{e^{-sk}(sk)^{j+1}}{(j-1)!s}\right]\\
     = & \frac{1}{\tau} \left[ \sum_{j=0}^{k-2}\frac{e^{-sk}(sk)^{j+1}}{j!s} -  \sum_{j=0}^{k-1}\frac{e^{-sk}(sk)^{j+1}}{j!s}\right]\\
     =&-\frac{s^{k-1}e^{-sk}k^k}{\tau (k-1)!} \leq 0
\end{align*}
for all $s\geq 0$, establishing the concavity. Also observe that \Cref{eq:1st_derivative} is positive for $s$ close to 0, so either the problem is unbounded, or first-order conditions will imply optimality. It is easy to see that the problem is not unbounded. This is because $\E(\min\{\poiss(sk),k\}) \leq k$ and $\tau<1$, so with $s$ large enough the objective function will become negative. We can conclude that the optimal value $s^*_\beta$ is the solution of making \Cref{eq:1st_derivative} equal to 0.

The form of $s_\alpha^*$ also comes from the fact that the objective function is concave. Thus, it will either satisfy the first-order condition in the $[0,1]$ interval, or it will be weakly increasing in the $[0,1]$ interval. In both cases the result holds. \qed

The form of $s^*_\alpha$ is implied by the concavity of the objective function combined with the constraint that $s\leq 1$.
% in $\alpha_k^\tau$.
We characterize the region where $s^*_\beta\leq 1$, in which $s^*_\alpha = s^*_\beta$ and our guarantee is tight.
\begin{proposition}\label{prop:simul_tight}
For any $k\in \mathbb{N}$ and $\tau \in (0,1)$ we have we have $s^*_\alpha = s^*_\beta$ if and only if
\[ \tau \leq 1 - e^{-k}\sum_{j=0}^{k-1} \frac{k^j}{j!}. \]
This condition holds for all $k$ if $\tau\leq 1/2$.
\end{proposition}
\proof{Proof.} \Cref{prop:simul_foc} implies that $s_\alpha^* = s_\beta^*$ if and only if $s_\beta^*\leq 1$. We will show that the condition of \Cref{prop:simul_tight} is equivalent to $s_\beta^*\leq 1$.

To show this we again resort to the concavity of the objective function in $\beta_{k}^\tau$. Since the function is concave, we have that $s_\beta^*\leq 1$ if and only if the derivative in \Cref{eq:1st_derivative} non-positive at $s=1$. Indeed, if it is non-positive, concavity implies that the first-order condition is satisfied and some $s'\leq 1$. On the other hand, if the derivative is positive at $s=1$, then the first-order condition is satisfied at some $s'>1$ and $s_\beta^* > s_\alpha^*$. This establishes the first statement of the lemma.

For the second statement, we see that
\[ 1 - \sum_{j=0}^{k-1} \frac{e^{-k}k^j}{j!} = \PP(\poiss(k)\geq k) \geq  \frac{1}{2}, \]
since $k$ is an integer \citep[Exercise~5.14]{mitzenmacher2017probability}. \qed

\begin{figure}[h]

    \centering
     \begin{subfigure}[b]{0.45\textwidth}
         \centering
         \includegraphics[width=\textwidth]{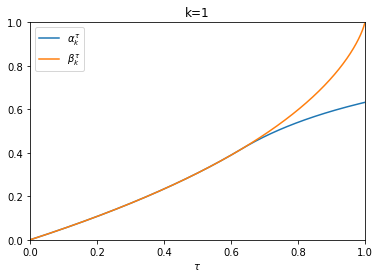}
     \end{subfigure}
     \hfill
     \centering
     \begin{subfigure}[b]{0.45\textwidth}
         \centering
         \includegraphics[width=\textwidth]{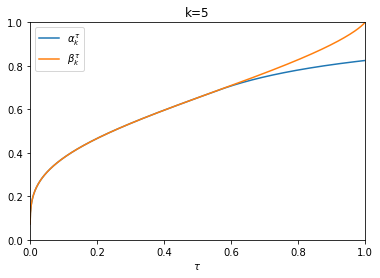}
     \end{subfigure}
     \hfill
     \centering
     \begin{subfigure}[b]{0.45\textwidth}
         \centering
         \includegraphics[width=\textwidth]{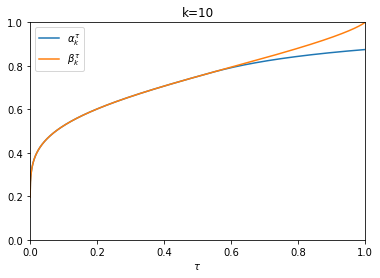}
     \end{subfigure}
     \hfill
     \begin{subfigure}[b]{0.45\textwidth}
         \centering
         \includegraphics[width=\textwidth]{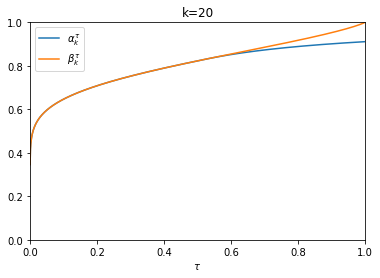}
     \end{subfigure}
     \hfill
        \caption{Values of $\alpha_k^\tau$ and $\beta_k^\tau$ for $\tau\in(0,1)$ and $k\in\{1,5,10,20\}$.}
        \label{fig:simul}
\end{figure}
\Cref{fig:simul} gives a graphic representation of $\alpha_k^\tau$ and $\beta_k^\tau$ for several values of $k$ and all $\tau\in(0,1)$. As stated in \cref{prop:simul_tight}, it can be shown that for all values of $\tau<1/2$, the bound is tight. It can also be observed that as $k$ grows, the guarantee becomes better, and that the gap between the upper and lower bounds becomes smaller.

\section{Numerical Study \label{sec:numerics}}

In this section, we provide numerical experiments that complement our theory and allow us to compare the rewards obtained under the three different models of hiring: $\seq$, $\paral$ and $\simul$.

Our experimental setting is based on the one by \citet{purohit2019hiring}. In our experiments, we sample 50 randomly generated pools of $n=100$ candidates. As in \citet{purohit2019hiring}, we generate candidates such that there is a \textit{negative correlation} between each candidate's value and their probability of acceptance, motivated by there being more market competition for the high-valued candidates. We also provide results where the sampling of values and acceptance probabilities is independent.

\orrev{The candidate pools are sampled in the following way. For each candidate $i$, $v_i$ is sampled independently from a Uniform(0,1) distribution. In the `negative correlation' setting the acceptance probability of candidate $i$, $p_i$, is sampled from a Beta$(10(1-v_i), 10v_i)$ distribution. In the `independent' setting, $p_i$ is sampled from a Uniform(0,1) distribution, independent of $v_i$. This very specific sampling procedure is the one used in \cite{purohit2019hiring}, \orrev{and as such we acknowledge that the insights we obtain in these experiments do not necessarily hold in general}. We choose to sample 50 candidate pools and average out the performance of the different heuristics and benchmarks across these pools. This helps to smooth out idiosyncratic choices made by algorithms in specific instances that may be misleading.}

%For details about the sampling of candidate pools, see \Cref{ap:num_sampling}.  in \Cref{ap:num_indep}.

\subsection{Implemented Policies \label{sec:exppolicies}}

We implement four heuristics for $\seq$ and one heuristic for $\paral$. The first sequential heuristic is $\ALG_\seq'$, a slight variant of $\ALG_\seq$ which is de-randomized, and in the case that the original algorithm would send less than $T$ offers, it fills those gaps with candidates with high $v_i$. The second heuristic was introduced by \citet{purohit2019hiring} and we call it `Adaptive sequential'. It corresponds to the optimal \textit{adaptive} algorithm that \textit{sends offers in decreasing order of $v_i$}. Since any non-adaptive algorithm can be improved by sending offers in decreasing order of $v_i$, Adaptive sequential serves as an upper bound for the performance of any non-adaptive algorithm, including our $\ALG_\seq$. The other two heuristics that we implement for $\seq$ are what we call the naive non-adaptive policies: `Value-ordered' ($\VO_\seq$) and `Expected-value-ordered' ($\EO_\seq$). These policies greedily send offers to the next candidate with the highest $v_i$ and $v_ip_i$, respectively.
%send offers to the $T$ candidates with the highest $v_i$ and $v_ip_i$, respectively. \wnote{still sorting by decreasing value of $v_i$?}\bnote{The second one does not sort by $v_i$, which I now realize is kind of stupid. However, I think that \cite{purohit2019hiring} also don't sort by $v_i$}.
For reference, we include the value of $\LP_\seq$. For $\paral$ we implement $\ALG_\paral'$, a de-randomized heuristic based on $\ALG_\paral$.

For $\simul$ we implement three heuristics. The first one is `Value-ordered' ($\VO_\simul$), corresponding to the optimal value-ordered policy that sends offers to the first $m$ candidates with the highest $v_i$ (and $m$ is optimized). Similarly, `Expected-value-ordered' ($\EO_\simul$) corresponds to the optimal policy that sends offers to the first $m$ candidates with the highest $v_i p_i$ (and $m$ is optimized). We also implement a greedy heuristic and include the value of $\LP_\simul$ for reference. 

\orrev{A detailed description of the implementation of these policies can be found in \Cref{ap:num_algo_description}.}
%For implementation details of all heuristics, see \Cref{ap:num_algo_description}.

\begin{figure}[h]
    \centering
    \includegraphics[scale = 0.51]{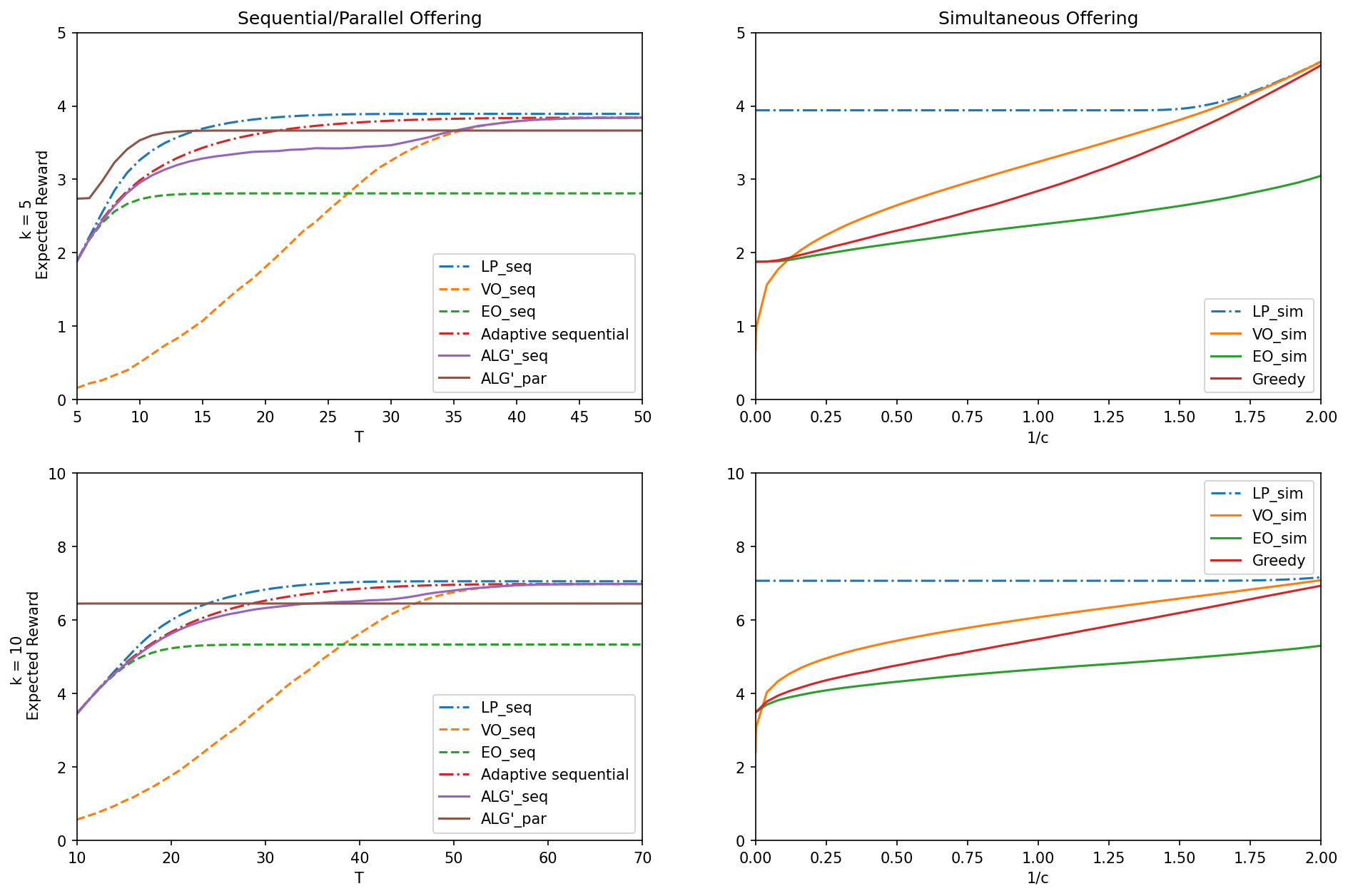}
    \caption[]{Numerical results for negatively correlated $v_i$ and $p_i$, averaged over the 50 random candidate pools. }
    \label{fig:numerics_negative_corr}
\end{figure}

\subsection{Results for the Negative Correlation Setting}
\Cref{fig:numerics_negative_corr} shows the results obtained for the implemented heuristics. \orrev{The left column shows how the expected reward changes as the number of offering rounds $T$ increases in the sequential and parallel offering settings. The right column shows how the expected reward changes as $1/c$ varies in the simultaneous setting. Recall that $c$ is the penalty for hiring each candidate over capacity $k$. We make the $x$-axis $1/c$ instead of $c$ so that the plots are more comparable to the sequential/parallel setting. The top row shows the results for $k=5$ positions and the bottom row shows the results for $k=10$ positions.}
% In what follows we provide several insights from the experiment.

\subsubsection{Insights about Sequential and Parallel Offering.}

The plots for $\seq$ and $\paral$ show how the expected reward of different heuristics grows as $T$ increases.
% grows and we are allowed to send offers to more candidates.

\textit{Difficult values of $T$ for sequential hiring.} A first observation is that intermediate values of $T$ are harder to approximate, both for the adaptive heuristic and the non-adaptive heuristics. This can be concluded by observing a higher gap between the heuristics and the benchmark upper bounds. It comes with no surprise, since extreme values of $T$ have trivial solutions. On one hand, if $T=n$, then the optimal policy is to send offers to all candidates in decreasing order of $v_i$. If $T=k$, then the optimal policy is to send offers to the $k$ candidates with the highest $v_i p_i$. Consequently, $\VO_\seq$ is (near-)optimal when $T$ is large, and $\EO_\seq$ is (near-)optimal when $T$ is small. The intermediate values produce more separation in the performance of different algorithms.

\textit{Virtue of our non-adaptive sequential policy.} We observe that $\ALG_\seq'$ outperforms both naive policies for all values of $T$. Given the constraints in $\LP_\seq$, our heuristic $\ALG_\seq'$ will behave like Expected-value-ordered when $T=k$.
%, \sout{and it will take more and more risk as $T$ grows} \wnote{What does "taking more risk" mean?  I would probably just remove this phrase.}.
There is a value of $T$ where the constraint $\sum_{i\in[n]} y_i \leq T$ stops binding, so the LP solution would remain constant for higher values of $T$, and the output of our algorithm would make less than $T$ offers. At this point, $\ALG_\seq'$ includes candidates in decreasing order of $v_i$ until $T$ offers are to be sent, and the policy starts behaving like $\VO_\seq$. This non-adaptive policy based on our LP correctly transitions from $\EO_\seq$ to $\VO_\seq$ as $T$ grows, and strictly dominates them on the difficult, intermediate values of $T$.
% seems to achieve `the best of both worlds' parsimoniously transitioning from conservative to risky as the amount of offers we are allowed to send grows.

\textit{More positions help the heuristics.} As $k$ becomes larger, our heuristics achieve better performance with respect to the available benchmarks: a smaller gap can be observed between $\ALG_\seq'$ and either the Adaptive heuristic or LP upper bound. A smaller gap can also be observed between the Adaptive heuristic and the LP upper bound. This is consistent with our theoretical results, where the approximation guarantee of our algorithm grows with $k$.

\textit{Parallel vs. Sequential Offering.} There is value in the ability to send more offers by sending them in parallel, and this value increases with $k$. This is natural since parallel algorithms are allowed to send as many offers as positions available, so the more available positions in the beginning, the more potential offers there are to be sent. In fact, when $k$ is small (there are two or three positions) the firm can gain more value by using a sequential and adaptive policy than by using a parallel, non-adaptive policy, even with the same amount of allowed offering rounds $T$. There is a point where $T$ becomes large enough such that the Adaptive heuristic, $\ALG_\seq'$ and $\VO_\seq$ outperform our parallel heuristic. This is because our parallel heuristic balances out the good candidates along $k$ lists, so when a candidate accepts (and no other candidate in their list receives an offer) many good candidates are left out. With $T$ large enough, the sequential policies (except $\EO_\seq$) are hedged against this because they have enough time to send offers to all the good candidates. An interesting direction is to design parallel algorithms that adaptively choose the next candidates.

\subsubsection{Insights about Simultaneous Offering.}

In the plots for $\simul$ we show how the expected reward of the implemented heuristics as $1/c$ grows.\footnote{The choice of showing the results with respect to $1/c$ instead of $c$ is because it allows showing large values of $c$ while keeping the scale of the graph reasonable. We can use $1/c \in (0,2]$ to concisely show $c\in[1/2,\infty)$, a range that allows us to compare the performance of $\VO_\simul$ against the performance of the sequential heuristics.}

\textit{Best policy depends on $c$.} For small values of $c$, $\VO_\simul$ outperforms Greedy and $\EO_\simul$, while for large values of $c$, the opposite occurs. This is natural since a very large value of $c$ is analogous to having a hard constraint of $k$ on the number of candidates hired. As $c$ becomes smaller, $\VO_\simul$ becomes better since there is room for error, and sending offers to candidates with high value with the risk of going over capacity becomes more valuable than hiring lower value candidates with a lower chance of not going over capacity. There is a value of $c$ low enough (namely $\min_{i\in[n]} v_i$) when all policies agree to send an offer to every candidate $i\in [n]$.

\textit{Larger $k$ helps Value-ordered.} We see that if $k=10$, $\VO_\simul$ outperforms the other two heuristics even for larger values of $c$. This can be explained by the fact that when $k=10$, there is more margin to capture all the high-value candidates without too much risk of going over capacity.

\textit{Simultaneous vs. Sequential Offering.} This experiment allows us to compare which values of $T$ and $c$ achieve a comparable expected reward. For instance, if $k=10$, $\ALG_\seq$ and $\VO_\seq$ with $T=30$ time steps achieve rewards comparable to $\VO_\simul$ and Greedy in the simultaneous setting with $c\approx 0.6$. A finer mapping from $T$ to $c$ can be done if one fixes a specific policy for each setting. Note that as $c \to \infty$, $\EO_\simul$ converges to the same value as $\EO_\seq$ when only $T=k$ offers can be sent, as having a large value of $c$ is equivalent to $k$ being a hard constraint.

%, for a fixed sequential and simultaneous policy,  For instance, the $k=5$ scenario with $\ALG_\seq$ for the sequential setting and Value-ordered for the simultaneous setting. A cost $c=1$ is roughly equivalent to having $T=13$ offers in $\seq$. Being able to send $T=10$ offers is roughly equivalent to a cost $c=1.32$. \wnote{Can you make this comparison on a point where it is not so sensitive to the specific Sequential policy chosen?  E.g. maybe $k=10$, $T=30$, would be similar to c=1/2, since most of the Sequential/Simultaneous policies would yield the same comparison} 
% Therefore, the optimal policy in the setting is to send offers to the $k$ candidates with the highest $v_i p_i$, which is also the optimal policy in the sequential setting when $T=k$.
%\vspace{-40pt}

\subsection{Results for the Independent Setting \label{ap:num_indep}}
\begin{figure}[h]
    \centering
    \includegraphics[scale = 0.51]{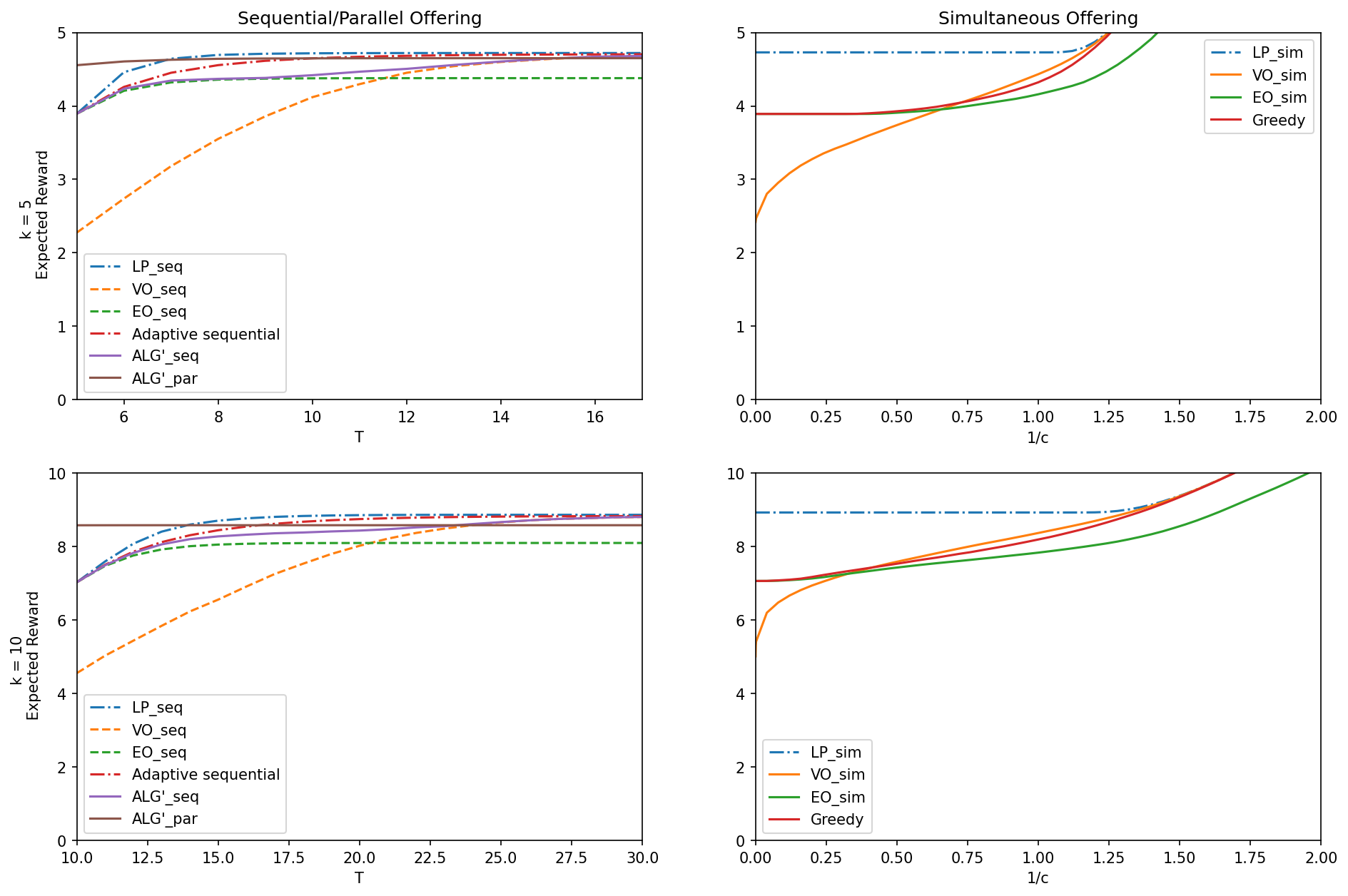}
    \caption[]{Numerical results for independent $v_i$ and $p_i$, averaged over the 50 random candidate pools. The left column shows how the expected reward changes as the number of offering rounds $T$ increases in the sequential and parallel offering settings. The right column shows how the expected reward changes as $1/c$ varies in the simultaneous setting.\protect\footnotemark The top row shows the results for $k=5$ positions and the bottom row shows the results for $k=10$ positions.}
    \label{fig:numerics_indep}
\end{figure}
% \footnotetext{Recall that $c$ is the penalty for hiring each candidate over capacity $k$.
% We make the $x$-axis $1/c$ instead of $c$ so that the plots are more comparable to the sequential/parallel setting.
% }
Figure \ref{fig:numerics_indep} shows the results obtained for the numerical experiments in the setting where $v_i$ and $p_i$ are independently generated (instead of negatively correlated).

Many of the insights of the negative correlation case also appear here: larger $k$ helps all heuristics, intermediate values of $T$ are the most difficult to approximate (sequential) and the optimal simultaneous policy depends on the value of $c$. However, some differences can be observed. These differences are mainly explained by the fact that the independent sampling of $p_i$ and $v_i$ generates more high-value, high-probability candidates, as opposed to the negative correlation case where fewer `unanimous' candidates --who will indisputably receive an offer-- appear.

\textit{Sequential - Less time required for optimality.} A smaller time horizon $T$ is required for $\ALG_\seq'$ and $\VO_\seq$ to achieve nearly the same performance as the upper bound benchmarks. This can be explained by the existence of more  high-value, high-probability candidates, who will fill up the positions and all policies will agree to send offers to them. This also explains the fact that $\ALG_\seq'$ only slightly outperforms $\EO_\seq$ for $k=5$ and small values of $T$.

\textit{Simultaneous - Value-ordered performs worse.} It can be seen that $\VO_\seq$ performs poorly with respect to the benchmarks, only outperforming Greedy in a slim range of $c$. This can again be explained by the existence of more  high-value, high-probability candidates. Since $\VO_\seq$ defines a threshold, in order to include all the high-value, high-probability candidates it must also include high-value, low-probability candidates, who will make the policy pay a higher penalty.

\subsection{Detailed Description of Heuristics \label{ap:num_algo_description}}

\subsubsection{Sequential Heuristics}
\begin{itemize}
    \item $\ALG_\seq'$: This heuristic is based on $\ALG_\seq$. It first solves $\LP_\seq$ and forms the two possible resulting lists of candidates that DR could have as an outcome. Call these lists $L_1$ and $L_2$. If any list $L_j$ contains less than $T$ candidates, we fill out the remaining slots with candidates from $[n]\setminus L_j$ in decreasing order of $v_i$. After both lists contain $T$ candidates, the policy evaluates the expected reward obtained by sending offers to the candidates of each list in decreasing order of $v_i$ and selects the list with the highest reward.
    
    \item Adaptive Sequential: This policy by \citet{purohit2019hiring} corresponds to the optimal adaptive policy, out of all the adaptive policies that send offers in decreasing order of $v_i$. Specifically, it solves the following dynamic program. Let $S(i,\ell,t)$ be the expected reward of hiring at most $\ell$ candidates in $t$ time steps by only considering candidates in $\{i,\dots,n\}$. The optimal value is given by $S(1,k,T)$ and the recursion solved is
    \[ S(i,\ell,t) = \max\{p_i(v_i + S(i+1,\ell-1, t-1) + (1-p_i)S(i+1, \ell, t-1), S(i+1, \ell, t)  ). \} \]
    
    \item Value-ordered ($\VO_\seq$): This policy greedily chooses the next candidate to send an offer to, selecting the one with the highest $v_i$ out of the remaining candidates.
    
    \item Expected-value-ordered ($\EO_\seq$): This policy greedily chooses the next candidate to send an offer to, selecting the one with the highest $v_i p_i$ out of the remaining candidates.
    
\end{itemize}

\subsubsection{Parallel Heuristic}

\begin{itemize}
    \item $\ALG_\paral'$: This policy first solves $\LP_\seq$ with $kT$ time periods instead of $T$. Two possible pools are constructed (the two possible outcomes of DR), and if they contain less than $kT$ candidates they are filled in the same fashion as in $\ALG_\seq'$. For each of the pools, we construct $k$ different lists as follows. Initiate the lists empty and initiate $M_j = 0$. For each of the candidates in the pool, in decreasing order of $v_i$, assign Candidate $i$ to the list $j$ that contains less than $T$ candidates and has the lowest $M_j$, and update $M_j \leftarrow M_j + p_i$. The idea is to spread out the high-value candidates among the lists while maintaining a balanced expected number of acceptances for each list. Once both sets of lists are constructed, the policy evaluates both of them and chooses the one with the highest expected value.
    
\end{itemize}

\subsubsection{Simultaneous Heuristics}

\begin{itemize}
    \item Value-ordered ($\VO_\simul$): Given that the candidates are labeled such that $v_1 \geq v_2 \geq \cdots \geq v_n$, this policy chooses the optimal $m \in\{1,\dots,n\}$ such that all candidates in $\{1,\dots,m\}$ receive offers.
    
    \item Expected-value-ordered ($\EO_\simul$): Given that the candidates are labeled such that $v_1 p_1 \geq v_2 p_2 \geq \cdots \geq v_n p_n$, this policy chooses the optimal $m \in\{1,\dots,n\}$ such that all candidates in $\{1,\dots,m\}$ receive offers.
    
    \item Greedy: This policy starts with an empty solution and iteratively adds the candidate that adds the highest marginal expected reward to the solution. It stops at the point that adding any candidate would reduce the expected reward obtained.

\end{itemize}

\end{APPENDICES}

%\theendnotes

% References here (outcomment the appropriate case)

% CASE 1: BiBTeX used to constantly update the references
%   (while the paper is being written).
%\bibliographystyle{informs2014} % outcomment this and next line in Case 1
%\bibliography{bibliography} % if more than one, comma separated

% CASE 2: BiBTeX used to generate mypaper.bbl (to be further fine tuned)
%\input{mypaper.bbl} % outcomment this line in Case 2

%If you don't use BiBTex, you can manually itemize references as shown below.

%%%%%%%%%%%%%%%%%
\end{document}